\documentclass[twocolumn,trackchanges]{aastex62} 
\usepackage{layouts}
\usepackage{amsmath}
\usepackage{rotating}
\usepackage{natbib}
\newcommand{\eg}{\textit{e.g.}}
\newcommand{\ie}{\textit{i.e.}}
\newcommand{\ps}{~s\(^{-1}\)}
\newcommand{\pss}{~s$^{-2}$}
\newcommand{\ha}{H$\alpha$}

\DeclareMathOperator*{\argmax}{argmax}

\newcolumntype{P}[1]{>{\centering\arraybackslash}p{#1}}
\newcolumntype{M}[1]{>{\centering\arraybackslash}m{#1}}
\newcolumntype{L}[1]{>{\raggedright\arraybackslash}m{#1}}

\newcommand{\newacronym}[4][]{\newcommand#2{#4 (#3\ifthenelse{\equal{#1}{}}{}{; \citealp{#1}})\gdef#2{#3}}}
\newcommand{\resetacronym}[4][]{\renewcommand#2{#4 (#3\ifthenelse{\equal{#1}{}}{}{; \citealp{#1}})\gdef#2{#3}}}

\newacronym{\euv}{EUV}{extreme ultraviolet}
\newacronym{\sdo}{SDO}{\textit{Solar Dynamics Observatory}}
\newacronym{\aia}{AIA}{\textit{Atmospheric Imaging Assembly}}
\newacronym[2012SoPh..275..207S]{\hmi}{HMI}{\textit{Helioseismic and Magnetic Imager}}
\newacronym[2008SSRv..136....5K]{\stereo}{STEREO}{\textit{Solar Terrestrial Relations Observatory}}
\newacronym[2004SPIE.5171..111W]{\euvi}	{EUVI}{\textit{Extreme Ultraviolet Imager}}
\newacronym[1995SoPh..162....1D]{\soho}{SOHO}{\textit{Solar and Heliospheric Observatory}}
\newacronym[1995SoPh..162..233H]{\cds}{CDS}{\textit{Coronal Diagnostic Spectrometer}}
\newacronym[1995SoPh..162..189W]{\sumer}{SUMER}{\textit{Solar Ultraviolet Measurements of Emitted Radiation}}
\newacronym[1995SoPh..162..357B]{\lasco}{LASCO}{\textit{Large Angle Spectroscopic Coronagraph}}
\newacronym[2007SoPh..243...19C]{\eis}{EIS}{\textit{EUV Imaging Spectrometer}}
\newacronym[2014SoPh..289.2733D]{\iris}{IRIS}{\textit{Interface Region Imaging Spectrograph}}

\begin{document}

\author[0000-0002-9601-1294]{Ivan P. Loboda}
\author[0000-0002-5448-8959]{Sergej A. Bogachev}
\affiliation{P.N. Lebedev Physical Institute of the Russian Academy of Sciences, 
53 Leninskiy Prospekt, 119991 Moscow, Russian Federation}
\correspondingauthor{Ivan Loboda}
\email{loboda@sci.lebedev.ru}

\title{What is a Macrospicule?}

\shorttitle{What is a macrospicule?} 
\shortauthors{Loboda \& Bogachev}

\begin{abstract} 
Macrospicules are typically described as solar jets that are larger and longer-lived than spicules, and visible mostly in transition-region spectral lines.
They show a broad variation in properties, which pose substantial difficulties for their identification, modelling, and the understanding of their role in the mass and energy balance of the solar atmosphere.
In this study, we focused on a sub-population of these jets that undergo parabolic trajectories when observed in the \ion{He}{2} 304~\AA{} line using high-cadence observations of the \aia{} on board the \sdo{} to accumulate a statistically significant sample, which included 330 such events.
We found these jets to be typically narrow (3--6~Mm), collimated flows of plasma, which reach heights of about 25~Mm, thus being among the smallest jets observed in the \euv{}.
Combined with the rise velocities of 70--140~km\ps{} and lifetimes of around 15~min, this makes them plausible candidates for the \euv{} counterpart of type~II spicules.
Moreover, we have found their dynamics to be inconsistent with a purely ballistic motion; instead, there is a strong correlation between the initial velocities and decelerations of the jets, which indicates that they may be driven by magneto-acoustic shocks with a dominant period of $10 \pm 2$~min. This makes these \euv{} jets similar in their dynamics to the conventional, or type~I spicules, thus justifying the name of macro-spicules in this case, while a substantial difference in the shock periods (1--2~min for the chromospheric jets) suggests a dissimilarity in the formation conditions.

\end{abstract}


\resetacronym{\euv}{EUV}{extreme ultraviolet}
\resetacronym[2012SoPh..275....3P]{\sdo}{SDO}{\textit{Solar Dynamics Observatory}}
\resetacronym[2012SoPh..275...17L]{\aia}{AIA}{\textit{Atmospheric Imaging Assembly}}

\section{Introduction} \label{sect:intro}

The term ``macrospicule'' was introduced by \citet{1975ApJ...197L.133B} in reference to the newly found jet-like phenomena observed in \textit{Skylab}'s \euv{} spectroheliograms at the solar limb.
The name refers to their morphological resemblance to conventional \ha{} spicules \citep{1945ApJ...101..136R, 1968SoPh....3..367B, 1972ARA&A..10...73B}, while being significantly larger in size and longer lived.
This and subsequent studies have shown macrospicules to reach heights from 7 to 70~Mm being 3 to 16~Mm in diameter, and to attain maximum velocities from 10 to 150~km~s$^{-1}$ with their lifetimes ranging from 3 to 45~min \citep{1975ApJ...197L.133B,  1976ApJ...203..528W, 1989SoPh..119...55D, 1994ApJ...431L..59K, 2002A&A...384..303P, 2015ApJ...808..135B, 2017ApJ...835...47K}.

Macrospicules are most often visible in emission in the transition-region spectral lines, such as \ion{He}{2} 304~\AA{}, \ion{N}{4} 765~\AA{}, and \ion{O}{5} 630~\AA{}, formed at temperatures of around $8 \times 10^4$~K, $1.4 \times 10^5$~K, and $2.5 \times 10^5$~K respectively.
However, while some authors have noted that macrospicules are absent in the hotter coronal lines \citep{1975ApJ...197L.133B, 2005A&A...438.1115X, 2009ApJ...704.1385S, 2011A&A...532L...1M}, others reported the observation of their  faint counterparts in, \eg{}, \ion{Ne}{8} 770~\AA{} and \ion{Mg}{9} 368~\AA{} spectral lines, formed at $6.3 \times 10^5$~K and $10^6$~K respectively, in which respect the observed features closely border the hotter and larger coronal jets \citep{1997SoPh..175..457P, 2000A&A...355.1152B, 2002A&A...384..303P, 2007AdSpR..40.1021P}.
At the same time, measurements of the radio brightness temperature in the 4.8~GHz range showed macrospicules to consist of a cool core at (4--8)$\times 10^3 $~K surrounded by a hotter shell at (1--2)$\times 10^5$~K \citep{1991ApJ...376L..25H}.
The above sets a rather vague upper temperature limit of $3\times 10^5$--$10^6$~K for this kind of solar jets \citep{1975ApJ...197L.133B, 2011A&A...532L...1M}.

Being first detected in polar coronal holes, macrospicules have long been thought to be specific to that particular kind of magnetic environment \citep{1975ApJ...197L.133B, 1976ApJ...203..528W, 1989SoPh..119...55D, 1994ApJ...431L..59K}.
However, more recent studies have shown macrospicules to also be present in the quiet-Sun regions, although they are significantly less numerous there \citep{1998ApJ...509..461W, 2015ApJ...808..135B, 2017ApJ...835...47K}.
Furthermore, while the on-disk counterparts of macrospicules have not been identified with full certainty, it was suggested that blinkers --- transient brightenings in the transition-region spectral lines --- are the most likely candidates \citep{1997SoPh..175..467H, 2005A&A...436L..43O, 2006A&A...452L..11M, 2009ApJ...704.1385S}.

Although no substantial solar cycle variations in the properties of macrospicules have been established so far, recent studies have detected the presence of quasi-biennial oscillations in their maximum lengths, areas, and average velocities \citep{2018AdSpR..61..611K, 2018ApJ...857..113K}.
In addition, spectroscopic observations of macrospicules have demonstrated alternating blue- and red-shifts, which were unanimously interpreted as a manifestation of their rotational motion \citep{1998SoPh..182..333P, 2000A&A...355.1152B, 2005A&A...438.1115X, 2009ApJ...704.1385S, 2011A&A...532L...1M}.
Finally, several studies noted the parabolic trajectories of these jets, with decelerations significantly lower than that due to gravity on the solar surface, which was explained by the combination of their ballistic motion and strong inclination along the line of sight \citep{1976ApJ...203..528W, 1994ApJ...431L..59K}.

Jets of a similar size, later also referred to as macrospicules, have been observed in the \ha{} spectral line \citep{1955epds.book.....W, 1979SoPh...61..283L, 1994A&A...281...95L, 2000SoPh..194...59Z}.
While some authors have found no correlation between features observed in \ha{} and the \euv{} transition-region lines \citep{1975SoPh...40...65M}, others have argued that \euv{} macrospicules are the hot sheaths of \ha{} macrospicules, in the same way that \euv{} spicules are the hot sheaths of their \ha{} cores \citep{1983ApJ...267L..65D, 1998A&A...334L..77B}, and furthermore, that they are closely associated with the X-ray bright points \citep{1977ApJ...218..286M}.
Subsequent observations have shown the likely presence of two sub-populations with different morphologies and dynamics, only half of which are seen in the \ion{He}{2} 304~\AA{} line  \citep{1998ApJ...509..461W, 2004ApJ...605..511Y, 2005ApJ...629..572Y}; it was therefore argued that some of the jets previously identified as \ha{} macrospicules were in fact mini-filament eruptions \citep{1986NASCP2442..369H, 2000ApJ...530.1071W}.
Similarly, \citet{1999A&A...341..610G} have proposed that macrospicules should be distinguished from the polar surges of \citet{1967ApJ...147.1131G}, which have a much more complex structure when observed in \ha{}.

On the other hand, macrospicules are closely bordered by the population of larger and hotter coronal jets \citep{2016SSRv..201....1R}. While the standard X-ray coronal jets of \citet{1992PASJ...44L.173S} typically have no or very weak counterparts in transition-region spectral lines, the so-called blowout jets, which have been recently identified by \citet{2010ApJ...720..757M}, most often show a well-pronounced cool component, which is best viewed in the \ion{He}{2} 304~\AA{} spectral line \citep{2013ApJ...769..134M}.
These cool counterparts, when observed separately in the \euv{}, are sometimes also referred to as macrospicules, as the boundary between these two groups of jets is not clearly established \citep{2010A&A...510L...1K, 2013ApJ...770L...3K, 2014ApJ...783...11A}. 

This confusion is partly fuelled by the lack of theoretical understanding of how the macrospicules form and evolve, and therefore, how they are distinguished, in terms of their mechanism, from other kinds of solar jets.
Although jets of similar size have been reproduced in one- and two-dimensional numerical simulations, most often driven by the velocity or pressure pulse in the upper chromosphere, the actual physical mechanism driving macrospicules remains unknown \citep{1982SoPh...81....9S, 1994ESASP.373..179A, 2011A&A...535A..58M}. 
In particular, it is still a matter of discussion whether macrospicules are just rare occurrences of oversized spicules, or whether they are produced by an essentially different mechanism \citep{1977ApJ...218..286M, 1985ApJ...290..359B, 1992PASJ...44..265S, 2000SoPh..196...79S, 2000A&A...360..351W, 2004ApJ...605..511Y}.

The above discussion illustrates that the term ``macrospicule'' is customarily used when referring to various solar jets observed in the transition-region spectral lines that are larger than spicules and smaller than the more energetic coronal jets, with both boundaries not being clearly defined.
Such categorisation is mostly based on the size and morphology of the jets, which is furthermore highly dependent on the spectral channel being used; consequently, this broad definition potentially comprises phenomena of different physical nature, which naturally results in a broadly distributed and sometimes contradictory set of observed characteristics. 
This, in turn, leads to the question as to what exactly should be called a macrospicule.

A few  attempts have been made recently to approach macrospicules from a statistical point of view \citep{2015ApJ...808..135B, 2015JApA...36..103G, 2017ApJ...835...47K} by taking advantage of long-term, high-cadence, and high-resolution observations in the \euv{} range offered by the \aia{} operating on board the \sdo{}. 
However, in these studies, a lot depend on the definition of a macrospicule.
While \citet{2015ApJ...808..135B} considered only jets that were no longer than 145~Mm, fell back on the solar surface, and whose footpoints were located exactly on the solar limb, \citet{2017ApJ...835...47K} defined macrospicules as ``thin'' jet phenomena shorter than 70~Mm, with a visible connection to the solar surface, and, most importantly, preceded by a brightening in their base. 

In this study, we alternatively concentrate on the subclass of \euv{} transition-region jets, which are characterised by the parabolic trajectories of their spires.
We viewed this rather specific dynamic behaviour as a characteristic feature that can potentially reveal the underlying dynamic mechanism, be it either the ballistic scenario proposed previously by, \eg{}, \citet{1994ApJ...431L..59K}, or any other.
Moreover, such selection allows the decelerations of the jets to be unambiguously defined, measured, and studied from a statistical point of view.
We should stress, however, that only a sub-population of the jets that are typically referred to as macrospicules was therefore examined, and that, consequently, the results obtained in this study should not be extended to all other jets falling under this term.

In Section~\ref{sect:data_methods} of the paper, we outline the data processing pipeline, including the observations, jet identification, data reduction, and parameter extraction. 
We continue in Section~\ref{sect:results} with an overview of the obtained results, including the distributions and pairwise correlations of the jets' parameters.
Finally, we speculate in Section~\ref{sect:discuss} on the possible implications of these results, discuss the main sources of error, and delineate further prospects for the study of ``macrospicules''.

\section{Data and methods} \label{sect:data_methods}

\subsection{Observations} \label{sect:observations}

For this study, we employed exclusively the \aia{} observations in the narrow-band channel centred at 304~\AA{}.
In the quiet Sun, the main contribution to this channel is made by the resonant \ion{He}{2} spectral line excited at transition-region temperatures of $4 \times 10^4$ -- $2 \times 10^5$~K, with the temperature of maximum emission at $8 \times 10^4$~K \citep{1975MNRAS.170..429J, 1978SoPh...58..299C, 2000SoPh..195...45T}.
This clearly delineates the thin interface region between the cool, dense plasma of a jet, such as the classical macrospicule of \citet{1975ApJ...197L.133B}, and the hot corona, revealing it in this channel as a bright, luminous feature.
The downside to the strength of the \ion{He}{2} 304~\AA{} line is, however, the high optical thickness of the plasma, which allows only the outer sheath of the jet to be effectively observed, thus hampering the examination of its inner structure and dynamics.

To the benefit of the statistical approach, \aia{} offers a continuous set of full-disk observations which now cover a sizeable period of more than seven years, typically with a high cadence of 12~s and a correspondingly high angular resolution of 0.6\arcsec{}, as yet unsurpassed by other long-duration missions.
The dataset used for this study consisted of the fifteen 6-hour long sections, each of which ensure the best cadence available, spanning the six years from June 2010 to June 2015, with a total duration of 90 hours.
The dates, as well as the start and end times of the sections, are given in Table~\ref{tbl:obs_data}.
Most of these observations, however, were taken in the years 2010 and 2011, partly because of the increased large-scale solar activity closer to the solar maximum in 2014, which effectively blocked smaller features from view, but mainly due to a considerable signal level drop in the 304~\AA{} channel during the later stages of \aia{}'s operation, probably resulting from the accumulation of volatile contamination on the  telescope's optics and detector \citep{2014SoPh..289.2377B}, which impaired the reliable identification of jets and extraction of their parameters.

\begin{table}[t]
\centering
\caption{Observed time periods and parts of the limb.} \label{tbl:obs_data}
\begin{tabular}{lM{3em}M{3em}M{5em}M{4em}}
\hline\hline
Date & Start (UT) & End (UT) & Parts of the limb\tablenotemark{a} & Number of jets\\
\hline
2010 Jun 1 & 12:00 & 18:00 & N, S, E, W & 56 \\
2010 Jun 4 & 00:00 & 06:00 & N, S, E, W & 40 \\
2010 Jun 4 & 11:00 & 17:00 & N, S, E, W & 30 \\
2010 Sep 16 & 09:00 & 15:00 & N & 11 \\
2010 Dec 1 & 00:00 & 06:00 & N, S, W & 49 \\
2011 Feb 26 & 00:00 & 06:00 & N, S & 25 \\
2011 Feb 26 & 06:00 & 12:00 & N, S, E & 21 \\
2011 May 31 & 00:00 & 06:00 & N, S & 25 \\
2012 Jun 4 & 00:00 & 06:00 & N, S & 11 \\
2012 Jun 4 & 12:00 & 18:00 & N, S & 8 \\
2013 Jun 4 & 12:00 & 18:00 & S & 14 \\
2014 Apr 1 & 12:00 & 18:00 & N, S & 28 \\
2014 Jun 4 & 00:00 & 06:00 & N, S & 16 \\
2015 Mar 11 & 00:00 & 06:00 & N & 2 \\
2015 Jun 4 & 12:00 & 18:00 & S & 1 \\
\hline
\end{tabular}
\tablenotetext{a}{Notation: N --- north, S --- south, E --- east, W --- west.}
\end{table}

\subsection{Data preprocessing}

To efficiently navigate the data and identify individual features therein, we first reorganized each section of the dataset into a three-dimensional data array, or a data cube.
We considered only the off-limb parts of the images that are free from large-scale solar activity, such as active regions or prominences; the approximate locations of the processed areas are given in Table~\ref{tbl:obs_data}.
Since these regions of interest are located along the curved edge of the solar disk, we then performed a polar transformation of the corresponding parts of the images to facilitate further data processing.
The resulting rectangular arrays were successively stacked into one or more data cubes (depending on the number of regions selected in a single image), enabling us to trace the temporal evolution of the off-limb structures.

This procedure can be formalised as $C[\phi,h,t_k] = I_k[x,y]$, where $C$ is the resulting  data cube, and $I_k$ is the $k$-th image in the dataset.
The three dimensions of the data cube are correspondingly the polar angle $\phi$ counted clockwise with respect to the north pole, the height above the limb $h$, and the observation time $t_k$ of the image $I_k$.
The polar angle $\phi$ is thus somewhat equivalent to latitude, but changes linearly from 0 to 360\degr{}, which is convenient for the global positioning of the jets.
On the local scale, however, it proved more useful to change to the circumferential distance along the limb $l = R_\sun \phi$ (where $R_\odot$ is the solar radius), which is expressed in units of length and serves as the horizontal axis in the polar-transformed images.
The Cartesian coordinates of the image, $x$ and $y$, are related to the polar coordinates $\phi$ and $h$ of the data cube by the standard equations for the polar transformation.

The transformation itself was implemented using the cubic convolution interpolation method \citep{keys1981cubic, PARK1983258}.
To test the accuracy of the interpolation, we obtained the error statistics by comparing the result of consecutive direct and reverse transformations to the original image over a small subset of the data.
We thereby inferred the optimal value of the interpolation parameter to be $c=-0.8$ for the solar images used (while a value of $c=-0.5$ is recommended for the general-purpose imagery), which resulted in a mean absolute error not exceeding 0.59 CCD counts in 2010, and 0.33 in 2015, at which time the signal level was significantly lower, with a mean relative error lower than 2.7~\% for all observation periods.

Since the resolution of \aia{} images is 0.6\arcsec{}, and therefore the corresponding pixel scale in the source images is $\delta x = \delta y \simeq 435$~km, we chose the grid step of the data cube to be $\delta h = \delta l = 500$~km in both spatial dimensions to preserve a unity aspect ratio on the limb, with $\delta l$ corresponding to $\delta\phi = \delta l / R_\sun \simeq 0.041\degr$.
Since we considered only the off-limb parts of the images, the lower height boundary was set to 0 and the upper one to 80~Mm, almost double the highest jet studied here, to ensure that all of the jet material is registered. 
Finally, since the exposure time was constant at 2.9~s in the 304~\AA{} channel of \aia{}, no intensity normalisation was necessary in this study.

\subsection{Jet identification} \label{sect:jet_id}

\begin{figure*}[t]
\centering
\includegraphics{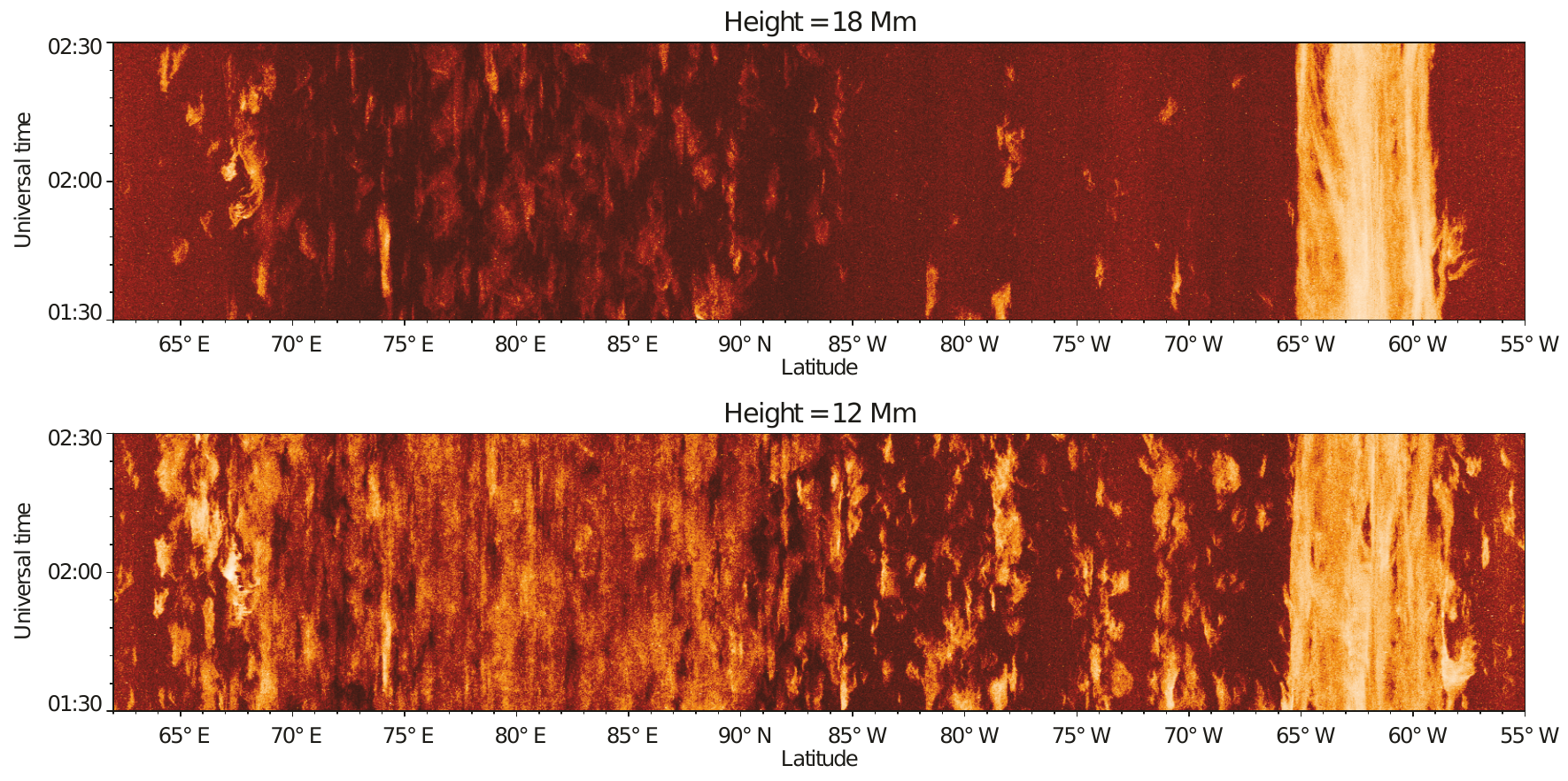}
\caption{Off-limb synoptic maps at different heights used for identification of the jets.}\label{fig:synoptic}
\end{figure*}

To effectively identify jets in these data, we examined the cross-sections $S_\mathrm{h}[\phi,t] = C[\phi,h_\mathrm{s},t]$ of each data cube at varied heights $h_\mathrm{s}$.
These cross-sections are in effect the synoptic maps, which reveal the off-limb transient structures with exceptional clarity: most such structures are visible as areas of enhanced brightness, typically with sharp borders due to the small thickness of the interface region.
Examples of such maps at two different heights are given in Fig.~\ref{fig:synoptic}.

At the lower heights (typically below $\sim 10$~Mm), which are dominated by the ``forest'' of \euv{} spicules \citep{1983ApJ...267..825W}, practically no individual structures can be discerned.
When the cross-section height $h_s$ is increased to about 10--15~Mm in the quiet Sun, and to 15--20~Mm in coronal holes, where the \euv{} spicules are noticeably longer, larger structures become visible as separate features in these maps.
At the same time, the heights above $\sim 30$~Mm are reached only in a small number of events and correspondingly are not suitable for jet identification.
We should note that, in most cases, jet bases were not observed being obscured by the optically thick \euv{} spicule ``forest''.
This results in uncertainties regarding a jet's position on the limb, as well as its length, lifetime, and initial velocity, which effect is discussed in more detail in Section~\ref{sect:disc_err}.

In these synoptic maps, the jets of primary interest typically show up as small, straight, narrow stripes.
However, this should not be the sole means of jet identification, and further investigation is necessary.
There are also several types of larger jets present, such as the blowout jets described by \citet{2010ApJ...720..757M, 2013ApJ...769..134M}, which show a much more complex structure and dynamics, and whose twisting motions are visible in these maps as a complex combination of curved bands.
Finally, the synoptic maps show a large number of features with no pronounced jet-like behaviour, which were excluded from this study.

We also used these synoptic maps for the identification of coronal holes and quiet-Sun regions --- the former being noticeably darker and populated with longer \euv{} spicules.
For example, in the synoptic maps shown in Fig.~\ref{fig:synoptic}, the coronal hole spans from approximately 67\degr{}~E to 84\degr{}~W.
In addition, a small portion of the limb between the 65\degr{} and 59\degr{} in the western hemisphere is dominated by a prominence, which appears as a large area of enhanced brightness.
The rest of the limb was therefore marked as quiet Sun.

\subsection{Data reduction} \label{sect:data_reduction}

\begin{figure*}[t]
\centering
\includegraphics{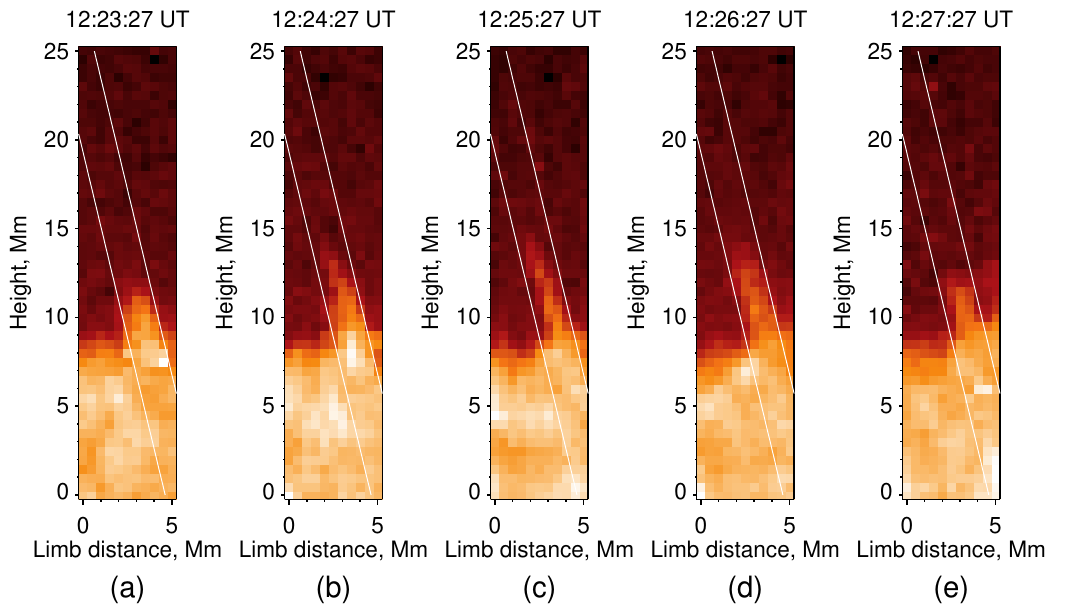}
\includegraphics{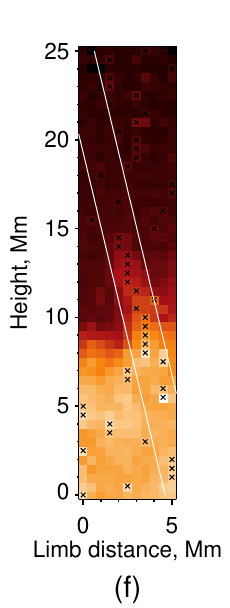}
\includegraphics{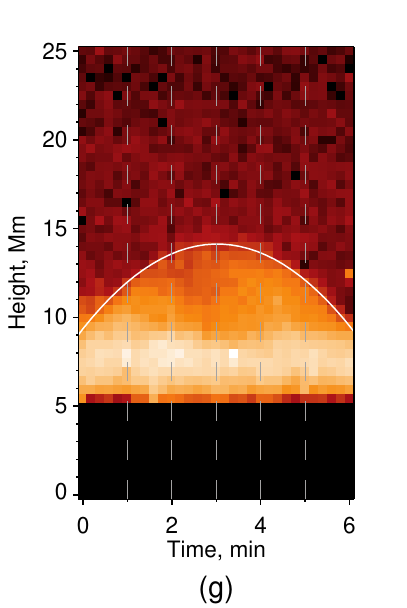}
\caption{Snapshots of a jet's evolution at selected timestamps (a)-(e), maximum intensity map (f), and time--distance diagram (g). The wide slit is indicated in panels (a)-(f) with white lines, the spine of the jet is shown in panel (f) with black crosses, timestamps of panels (a)-(e) are indicated in panel (g) with grey dashed lines, and the fitted parabolic trajectory is shown in panel (g) with the white curve.}
\label{fig:proc_steps}
\end{figure*}

Having identified a jet on the synoptic map, we isolated a small portion, $C'$, of the data cube, $C$, containing the jet itself and its immediate surroundings, thus substantially narrowing the cube dimensions both along the limb and in the temporal domain, but preserving its full range in the vertical direction.
For the preliminary classification of the jet, we examined the animated image sequence from $C'$, snapshots of which are shown in Figs.~\ref{fig:proc_steps}a-e.
Herein, the presence of both elevation and retraction phases was a necessary condition, although moderate fading of the jet material was acceptable.
Nevertheless, an earlier study of macrospicules indicated that most of these jets do not fade completely in the 304~\AA{} channel, despite losing up to 80~\% of their visible material \citep{2017A&A...597A..78L}.

Further processing steps mainly followed the data reduction method as described in \citet{2017A&A...597A..78L}, although with several modifications.
First, to determine the position of the jet's axis, a maximum  intensity map $$ M[l,h] = \displaystyle\max_{t}C' $$ was built, an example of which is shown in Fig.~\ref{fig:proc_steps}f.
In the case of no external interference, this map clearly shows the area covered by the jet in the course of its observed motion.
Subsequently, the spine of the jet was determined as a series of intensity maxima positions $$ l_\mathrm{s}[h] = \displaystyle\argmax_{l} M$$ calculated for every horizontal slab of the map, marked in Fig.~\ref{fig:proc_steps}f with dark crosses.
For a straight jet, a linear fit through the meaningful part of this spine was regarded as the jet axis. 
The lower and upper boundaries of the fitted region of the spine were thus automatically adjusted based on the normalised goodness of the fit and then manually controlled to ensure the exclusion of possible outliers in the spine.
This procedure yielded the inclination of the jet and the position of its assumed base on the limb, which is otherwise obscured by the \euv{} spicule ``forest''.

In this study, we neglected the internal structure of the jets, thus regarding them as unidirectional flows of plasma, and studied only their axial motions.
This approach was later justified by the fact that most of the jets studied were straight, unidirectional flows of plasma characterised by a simple morphology.
One of the most popular techniques in this approach is to take intensities from a narrow slit along the jet's axis, moving it, if necessary, in the perpendicular direction to follow the possible displacements of the jet.
The results, however, can be rather misleading due to the transverse plasma motions with respect to the slit that have not been accounted for \citep[for a discussion, see, \eg{},][]{2012ApJ...750...16Z}.

As an alternative, we consider a wide slit parallel to the jet's axis that covers the jet entirely in the maximum intensity map $M$, thus ensuring that all of its material is registered.
A counter-effect of this approach is, however, a lower signal-to-noise ratio, as more of the surroundings are thereby included with the relevant data.
Another limitation is that using this method it can be more difficult, if not even impossible, to isolate closely spaced jets, although in our case this was a fairly rare occurrence.
The width of this slit was taken as the maximum width of the jet --- one of the basic measured parameters  discussed among other results in Section~\ref{sect:results}.

Summing up the intensities in the horizontal direction within the wide slit, we obtained the time--distance diagram $W[t,h]$, which is a good visual representation of the axial plasma motions of the jet.
At this stage, jets can be classified based on their apparent dynamics, including the trajectory of the jet's spire in the image plane.
However, to define the latter reliably with an automated method, such as the dynamic thresholding technique described in Section~\ref{sect:fitting}, further processing of the time--distance diagram is required.

Being located close to the solar limb, the jets were observed against a relatively strong, but more importantly, inhomogeneous background, with a steep gradient in the radial direction \citep{1995ApJ...442..653J, 1999SoPh..188..259D}.
To determine this background, we relied on the fact that resonant scattering was shown to be the dominant mechanism for the \ion{He}{2} 304~\AA{} emission off limb \citep{2001A&A...380..323L}, which means that, in observations, the plasma structures are almost exclusively seen as bright features, and therefore, that in the absence of long-term activity, such as prominences and active regions whose lifetimes are comparable to or greater than the duration of the observation itself, a minimum signal $$ B[\phi,h] = \displaystyle\min_t C $$ is a good approximation to the background at a given point $(\phi,h)$ above the limb.
However, before calculating the minimum, we made sure to exclude all of the negative outbursts outlying the average signal by $3\sigma$ at that point, which could arise, for example, from data corruption.

For a particular jet, the background $B[\phi,h]$ was averaged inside the wide slit in the same way as for the time--distance diagram to provide a one-dimensional intensity profile $B_\mathrm{w}[h]$.
This profile was then smoothed by fitting it to a model function $\widetilde{B}(h)$, which we adopted from \citet{2015SoPh..290.1963L} and which proved to be a good representation of the quiet 304~\AA{} emission profile at heights greater than $10$~Mm above the limb.
Moreover, this function tends to infinity at the limb, which allows us to effectively suppress any transient structures lower than $5$~Mm.
The fully processed time--distance diagram was thus finally obtained as $W_\mathrm{b} =  \max\{W-\widetilde{B}, 0\}$, an example of which is shown in Fig.~\ref{fig:proc_steps}g.

\subsection{Trajectory fitting} \label{sect:fitting}

Given a certain threshold $T$, the spire of the jet can be defined as its highest point for which $W_\mathrm{b} > T$.
On the time--distance diagram, this gives the sequence of spire positions throughout the jet's lifetime, \ie{}, the jet's apparent trajectory $h_\mathrm{s}[t]$.
Since our selection is of jets with nearly parabolic motion, $h_\mathrm{s}(t)$ can be well approximated by the second-order polynomial $ h_{\mathrm{f}}(t) = p_0 + p_1 t + p_2 t^2 $, defined by the three coefficients $p_1$, $p_2$, and $p_3$.
The fitting algorithm used was a least-squares polynomial fit based on the matrix inversion.
Taking into account that the motion of the jet occupied most, but not all, of the time--distance diagram, the fitting was performed in a narrowed time interval $[t_1,t_2]$.

To evaluate the goodness of the fit, the relative error for the highest-order coefficient $\delta p_2 = \Delta p_2 / |p_2|$ was employed, where $\Delta p_2$ is the corresponding absolute error provided by the fitting algorithm.
Compared to the chi-squared statistic, this has the advantage of being a dimensionless value, and moreover, it does not depend on the choice of the origin of the coordinates.
Consequently, optimisation of this value was performed by varying the threshold $T$ in the range $[0, \frac{1}{2}\max\{W_\mathrm{b}\}]$ with a step of 0.5 CCD counts, and with fitted interval limits $t_1$ and $t_2$ restricted by the condition that $t_2 - t_1 > \frac{1}{2} \Delta t_{C'}$, where $\Delta t_{C'}$ is the temporal span of $C'$, which was manually, and thus arbitrarily, selected on the jet identification step (Section~\ref{sect:jet_id}).

From the optimal fit thus found, the major dynamic characteristics of the jet can be obtained: the maximum apparent height 
\begin{equation}
h_\mathrm{max} = p_0 - p_1^2 / 4 p_2 \,,
\label{eq:hmax}
\end{equation}
the jet's lifetime
\begin{equation}
t_\mathrm{life} = \sqrt{p_1^2 / p_2^2 - 4 p_0 / p_2} \,,
\label{eq:lifetime}
\end{equation}
the initial velocity
\begin{equation}
v_0^\ast = \sqrt{p_1^2 - 4 p_0 p_2} \,,
\label{eq:inivel}
\end{equation}
and the deceleration
\begin{equation}
a^\ast = -2 p_2 > 0 \,.
\label{eq:decel}
\end{equation}
Here, the lifetime and initial velocity are calculated by extrapolating the parabola down to the solar limb because of the unknown position of the jet's footpoints.
As a result, these values are subject to a measurement error, which is discussed in greater detail in Section~\ref{sect:disc_err}.
Also, due to the technique by which the time--distance diagrams were obtained, Equations~(\ref{eq:inivel}) and (\ref{eq:decel}) give only the vertical component of the velocity and deceleration.
Their two-dimensional components in the image plane, however, as well as the jet's visible length, could be easily reconstructed given the jet's apparent inclination angle.

\section{Results} \label{sect:results}

The method described above enables the fast identification of jets and the determination of their main spatial and dynamic characteristics with a minimum of manual operations. 
Within the observation period specified in Section~\ref{sect:observations}, we identified a total of 330 jets, with 209 (63.3~\%) found in the coronal holes and 121 (36.7~\%) in the quiet-Sun regions.
These are different magnetic environments, which could substantially alter the formation and evolution of the jets, and therefore we studied these two groups of jets independently, in addition to studying all of the jets combined.

\subsection{Birth rates}

Firstly, the bare fact of jet identifications, with their positions and times of occurrence known, provides us with their average birth rates.
To evaluate them as the total number of jets that would be produced per unit time across the entire surface of the Sun, we employed a simple geometric model as described in \citet{1998ApJ...509..461W}.
We have thus inferred an average birth rate of $1.1 \times 10^{-2}$~s$^{-1}$ considering both groups of jets, which is, however, noticeably higher for the coronal-hole jets at $2.3 \times 10^{-2}$~s$^{-1}$ and accordingly lower for the quiet-Sun jets at $7.2 \times 10^{-3}$~s$^{-1}$.
These numbers are close to those obtained by \citet{1998ApJ...509..461W} for the quiet-Sun \euv{} macrospicules ($0.02 \pm 0.02$~s$^{-1}$), but are about two orders of magnitude lower than for the same jets in the polar coronal holes; only the maximum birth rate, of around 2.0~s$^{-1}$, computed for the two most closely spaced jets in both spatial and temporal domains, is comparable to the birth rates of \citet{1998ApJ...509..461W} for the coronal-hole macrospicules (1--2~s$^{-1}$).
This is, however, not surprising given the fact that we only study a narrowly constrained subset of the \euv{} jets.
Moreover, our results are most probably observationally biased, since not all of the jets were taken into account, but rather only those that could be reliably processed by the algorithm described in Section~\ref{sect:data_methods}, meaning that the numbers we infer should be treated as a lower boundary.

We can also indirectly compare our results with those obtained in a number of previous works, from the durations of the observation periods used and the total numbers of jets identified.
Accordingly, both \citet{2015ApJ...808..135B} and \citet{2015JApA...36..103G} identified 101 macrospicules on the entire limb throughout 144 hours of observations, which amounts to roughly $8.7 \times 10^{-4}$~s$^{-1}$, while \citet{2017ApJ...835...47K} found 301 jets during a period of 576 hours, which yields a birth rate of around $6.5 \times 10^{-4}$~s$^{-1}$.
These much lower birth rates may, again, result from different definitions of a macrospicule, since the bright footpoint used in these studies for jet identification, if present, is most likely to be obscured by the optically thick \euv{} spicule ``forest''.
Moreover, the above estimates do not account for the fact that, in these works, certain parts of the limb were effectively blocked by large-scale solar activity, thus attenuating the calculated birth rates.
Similarly, the results of \citet{2017A&A...597A..78L} imply a much higher birth rate of up to 0.2~s$^{-1}$ for the coronal-hole macrospicules, since in this work all kinds of the jets were considered irrespective of their structure and dynamics, some of which were arguably polar surges rather than macrospicules.
However, this rate is still significantly lower than the results of \citet{1998ApJ...509..461W}.

\subsection{Parameter distributions} \label{sect:res_destribs}

\begin{figure*}[t]
\flushleft
\includegraphics{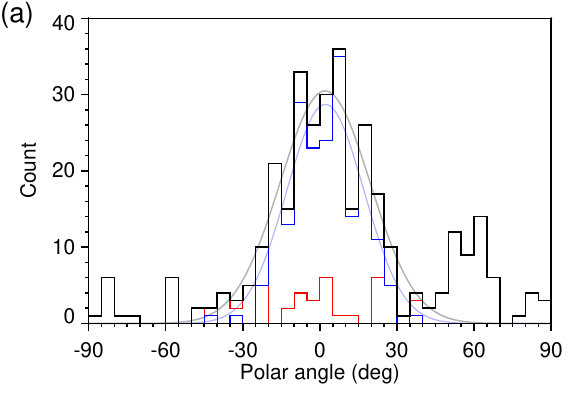}
\hfill
\includegraphics{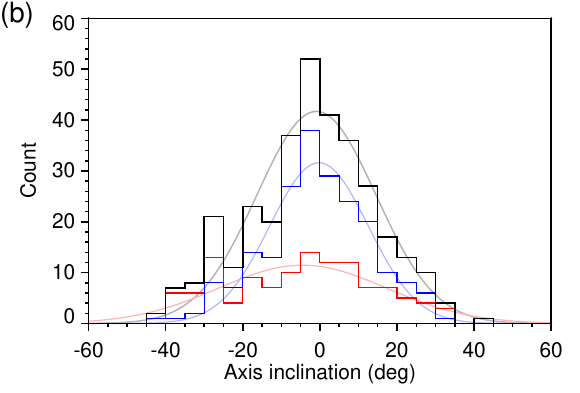}
\hfill
\includegraphics{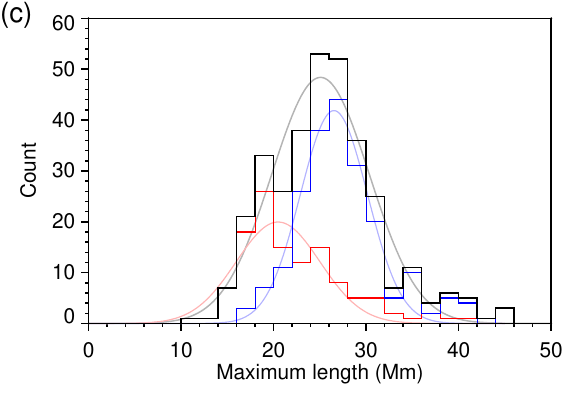}
\\
\includegraphics{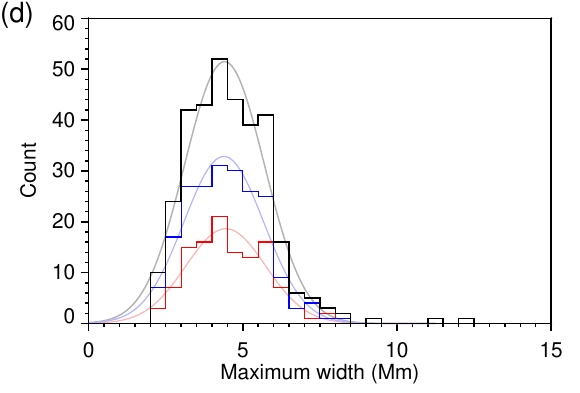}
\hfill
\includegraphics{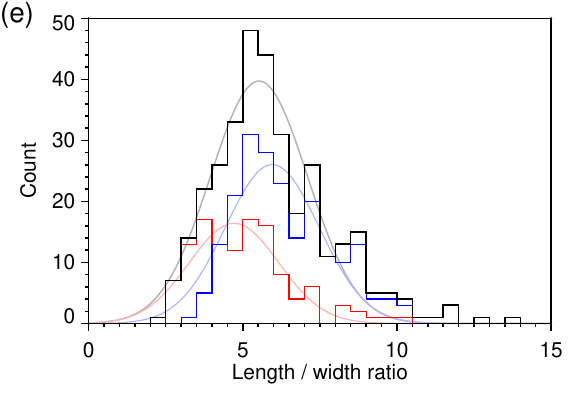}
\hfill
\includegraphics{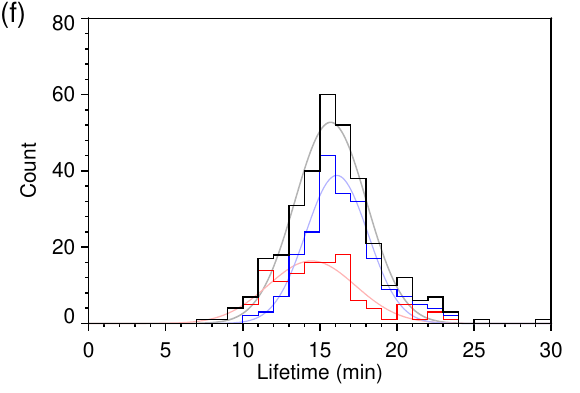}
\\
\includegraphics{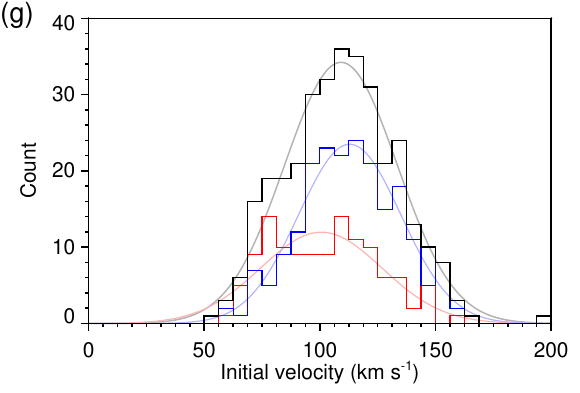}
\hfill
\includegraphics{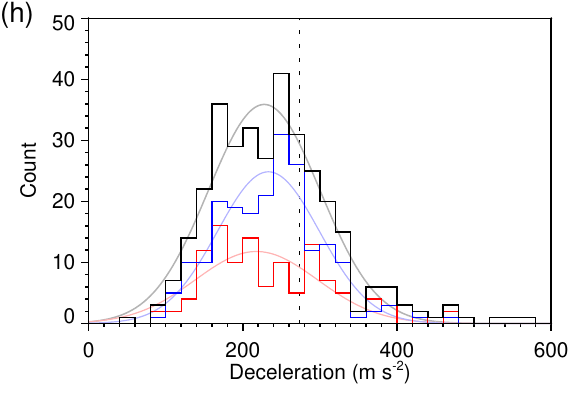}
\hfill
\includegraphics{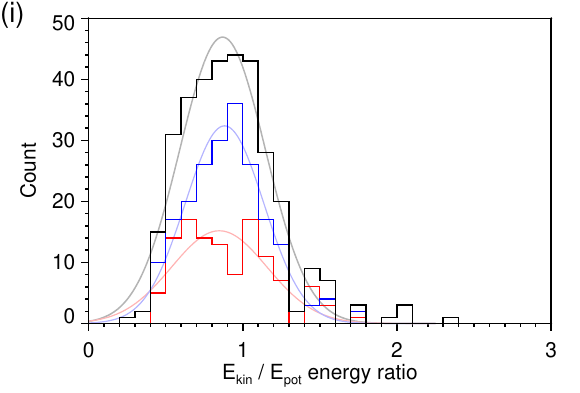}
\caption{Distributions of jet characteristics. Blue line is for coronal hole jets, red line is for quiet-Sun jets, and black line is for the two groups of jets combined. Light-red, light-blue, and grey curves are the respective Gaussian fits to these distributions. Dashed line in panel (h) denotes the free-fall acceleration due to gravity $g_{\odot} =274$~m\pss{} at the solar surface. Note that polar angle in panel (a) is relative to the nearest pole.}
\label{fig:distribs}
\end{figure*}

With the dynamics of the jets described by a series of numerical parameters, the properties of the whole population can be studied through the examination of their distributions and pairwise correlations.
Wherever possible, Gaussian fits to the obtained distributions were performed, which are plotted alongside the distribution histograms in Fig.~\ref{fig:distribs}.
In addition, a comprehensive set of numerical characteristics describing both the distributions (minimum and maximum values, mean, median, and standard deviation) and the fits (Gaussian mean and sigma) is given in Table~\ref{tbl:destr_par}. 

Most of the distributions obtained are unimodal, the only exception being the distribution of jet positions around the limb, which has two clear maxima at the solar poles.
We have thus found it useful to superimpose the north and south hemispheres on the same plot by redefining the polar angle as counted clockwise with respect to the nearest pole (Fig.~\ref{fig:distribs}a).
The resulting distribution thus has only one strong maximum, composed mostly of the coronal-hole jets, showing that they are concentrated around the poles despite the rising phase of the solar cycle, being located entirely within the range of $\pm 40\degr$.
On the contrary, the same distribution for quiet-Sun jets is significantly more uniform, except for a smaller maximum at around +60\degr{}, which we consider to be a fluctuation that would disappear given a larger and more regular dataset, and which was therefore not fitted with a Gaussian.

This contrasts with the results of \citet{2017ApJ...835...47K}, who have found quiet-Sun macrospicules to be positioned as a ``ring'' around the coronal holes, \ie{}, they have observed considerably fewer jets on the equator.
The disagreement is, in fact, even more pronounced, as our dataset was more focused at the poles and thus was more subject to observational bias.
It might, however, be that the full-limb observations of \citet{2017ApJ...835...47K} are biased as well, since the mid-latitudes are more likely to be blocked by larger forms of solar activity, such as prominences and active regions, on the rise of the solar cycle.
We can also compare the obtained distribution to the earlier results of \citet{2015JApA...36..103G}, who studied a set of 101 macrospicules defined loosely as jet-like features in the AIA's 304~\AA{} channel observations.
These authors fitted their distribution with a Gaussian and obtained a mean polar angle of 3.87\degr{} with the width of the distribution being $\sigma = 31.6$\degr{}, which is close to the standard deviation for the full set of jets in the present study.
However, when also fitted with a Gaussian, the distribution obtained here gives a substantially lower value of $\sigma = 17.4\degr$, which reflects the relatively narrow peak around the poles.
Additionally, both studies show only a small displacement of the distribution's mode from the poles, despite the fact that both observations were made during the rising phase of the solar cycle, when coronal holes are not necessarily centred around the poles.

The apparent inclinations of the jets (counted clockwise with respect to the normal) show nearly identical distributions for both groups of jets with no significant asymmetry (Fig.~\ref{fig:distribs}b).
In absolute values, the inclinations are relatively small, with 64.5~\% of jets having inclinations smaller than 15\degr{}, and with practically no jets inclined by more than $40\degr$.
This contradicts the assumption of \citet{1994ApJ...431L..59K} that inclinations of around 65\degr{} are necessary to explain the typical decelerations of \euv{} macrospicules.
For example, if a jet's inclination in the image plane is as high as 40\degr{}, the inclination along the line of sight still needs to be about 63\degr{} to satisfy this condition; from considerations of symmetry, however, one can expect the line-of-sight inclinations of the jets to be limited by $40\degr$ in either direction.
It is possible, however, that the most oblique jets remain unidentified as they are not sufficiently elevated above the \euv{} spicule ``forest'', which leads to a selection bias in the \ion{He}{2} 304~\AA{} observations.
On the other hand, our results are close to those of \citet{2009SoPh..260...59P} for \ha{} spicules, where inclinations were obtained in the range of $\pm 50\degr$, and to those of \citet{2012ApJ...759...18P} with inclinations mostly within $\pm 40\degr$ for the quiet-Sun and coronal-hole spicules observed in the \ion{Ca}{2}{ H} line.

The typical lengths of the jets are 16--32~Mm (for 86.1~\% of the sample), with coronal-hole jets being noticeably longer than the quiet-Sun jets by around 23~\% (Fig.~\ref{fig:distribs}b).
The mode of the distribution is well above the lower observable limit of 10--15~Mm, which allows us to conclude that the jets studied do not form a continual extension of the \euv{} spicules.
The numbers themselves are mostly in line with those of previous works (for a brief summary, see, \eg{}, Table~2 in \citealp{2017ApJ...835...47K}), although our results are most likely an underestimate due to the unknown position of the jet footpoints, which we expect to be located below the limb rather than above it; the further discussion of this effect is given in Sectoin~\ref{sect:disc_err}.
Nevertheless, \citet{2015ApJ...808..135B} and \citet{2017ApJ...835...47K} have both inferred the mean height of the jets to be 28.1~Mm, which is only 2.6~Mm higher than that measured herein. 
What is more important, however, is that these two studies have identified jets as long as 60.4~Mm and $\sim 65$~Mm respectively, while in the present study the maximum length was found to be 44.6~Mm despite the larger, or at least comparable, number of jets examined.
In our previous study, jets of a similar length were qualified as polar surges rather than macrospicules, having much more complex dynamics and consequently non-parabolic trajectories \citep{2017A&A...597A..78L}.

The maximum widths of the jets are comparatively small, with 79,1~\% falling into the range of 3--6~Mm (Fig.~\ref{fig:distribs}d).
Furthermore, these results should be treated as an upper limit due to the measurement technique used, which did not account for the possible transverse motions of the jets, their diffuse boundaries, and the variations of the jet's width throughout its lifetime or along the axis.
This being said, these numbers are among the smallest previously obtained, although widths in the earliest studies were typically close to the resolution limit of those observations \citep{1975ApJ...197L.133B, 1989SoPh..119...55D, 1994ApJ...431L..59K}.
Compared to the more recent results, the jets studied here are significantly narrower than those observed by \citet{2015ApJ...808..135B}, who found their mean width to be 7.6~Mm, and insignificantly wider than those in \citet{2017ApJ...835...47K}, even though the distribution obtained therein is more positively skewed, \ie{}, includes a larger fraction of wider jets.
We have also found no significant difference between the coronal-hole and quiet-Sun jets in this respect, which, given the significant difference in length, means that the jets in the coronal holes tend to be more elongated, but not generally larger as a whole.
The height-to-width ratios of the jets, the histogram for which is plotted in Fig.~\ref{fig:distribs}e, typically range from 4 to 8 (for 71.8~\% of the sample), in isolated cases reaching as high as 13.6, which supports our impression from the visual inspection of the jets that they are largely collimated, linear flows of plasma.

The typical lifetime of the jets is about 15~min, being between 13 and 18~min for 67~\% of jets (Fig.~\ref{fig:distribs}f), while their velocities are of the order of 100~km\ps{}, with 86.1~\% of jets having initial velocities from 70 to 140~km\ps{} (Fig.~\ref{fig:distribs}g).
The latter parameter is, therefore, mostly less than, or close to the typical sound speed in the lower corona, $c_\mathrm{s} = \sqrt{\gamma k_\mathrm{B} T/m} \simeq 150$~km\ps{}, where $\gamma = 5/3$ is the ratio of specific heats for the fully ionized coronal plasma, $k_\mathrm{B}$ is the Boltzmann constant, $T \simeq 10^6$~K is the temperature, and $m \simeq 10^{-27}$~kg is the mean particle mass \citep{2015ApJ...806...11M}.
With the exception of \citet{2015ApJ...808..135B}, who had one case of a macrospicule having a maximum velocity of 335~km\ps{}, which was therefore qualified as a probable outlier, both of these results are, again, close to those previously observed for the \euv{} macrospicules.
Finally, both lifetimes and initial velocities are slightly higher for the coronal-hole jets compared to the quiet-Sun jets by around 10~\%. 
This differs from the results of \citet{2015ApJ...808..135B}, who also found higher velocities for the coronal-hole macrospicules, but similar lifetimes irrespective of the jets' magnetic environment.

The most interesting results, however, relate to the jets' decelerations.
They show a broad distribution, with 73.6~\% of jets having decelerations within 140--300~m\pss{}, and with no significant difference between the jets located in coronal holes and in the quiet-Sun regions (Fig.~\ref{fig:distribs}h).
What is more important, however, is that a considerable fraction of jets --- 94, or 28.5~\% of the total number --- have decelerations larger than the free-fall acceleration due to gravity at the solar surface, $g_{\odot} =274$~m\pss{}, with the maximum detected deceleration being as high as $563 \pm 21$~m\pss{}, which certainly does not fit the ballistic scenario of jet dynamics.
On the other hand, the minimum deceleration measured is just $56.1 \pm 1.4$~m\pss{}, which implies a strong inclination of $78.2 \pm 0.3$\degr{} for this particular jet, which we do not view as probable since all of the jets examined had apparent inclinations not exceeding 40\degr{}.

Perhaps, a better way to assess the degree to which the dynamics of a jet is governed by solar gravity is to compare the kinetic energy at the onset of its motion and the potential energy at the jet's maximum elevation.
While the ratio $\xi = 1$ corresponds to a purely ballistic motion, the loss of energy would imply values of $\xi > 1$, and conversely, $\xi < 1$ indicates that the jet's material experiences additional upward acceleration, which is often referred to as the ``driving force", thus reaching heights higher than it would have been expected from its initial velocity.
Taking into account Equations~(\ref{eq:hmax}), (\ref{eq:inivel}), and (\ref{eq:decel}), and noting that $v_0^\ast = v_0\cos\alpha$ and $a^\ast = a \cos\alpha$, where $v_0$ and $a$ are the jet's initial velocity and deceleration in the image plane, and $\alpha$ is the jet's apparent inclination relative to the radial direction, we obtain
\begin{equation}
\xi = \frac{E_\mathrm{kin}}{E_\mathrm{pot}} = \frac{v_0^2}{2g_{\odot}h_\mathrm{max}} = \frac{a}{g_{\odot}\cos\alpha} \,,
\label{eq:en_rat}
\end{equation}
which also shows that this approach is equivalent to comparing the jets' decelerations in the image plane with the projection of the gravitational acceleration on their apparent axes $g_{\odot}\cos\alpha$.
From this perspective, as many as 117 jets, or 35.5~\% of the sample, have decelerations larger than solar gravity would imply, which corresponds to $\xi > 1$ (Fig.~\ref{fig:distribs}i).

\subsection{Correlations}

\begin{figure*}[t]
\flushleft
\includegraphics{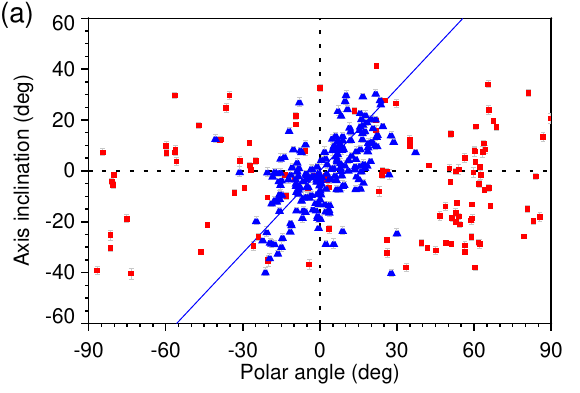}
\hfill
\includegraphics{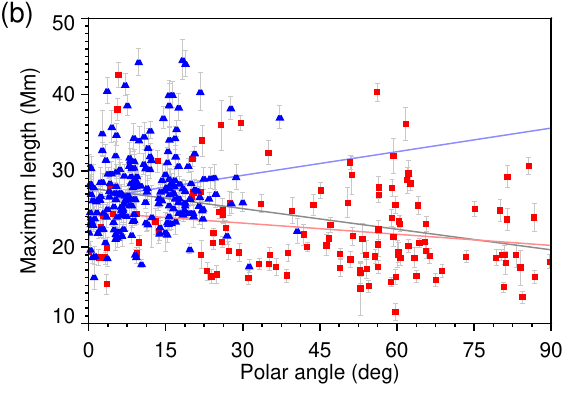}
\hfill
\includegraphics{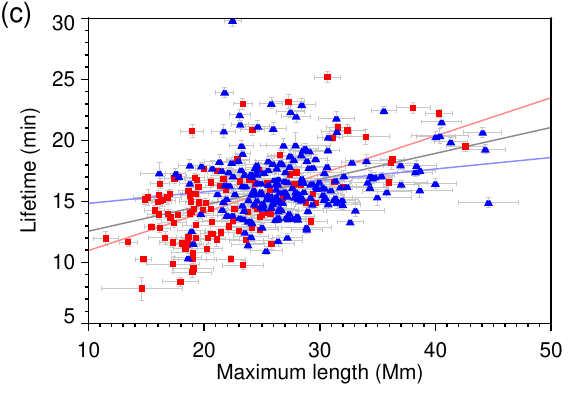}
\\
\includegraphics{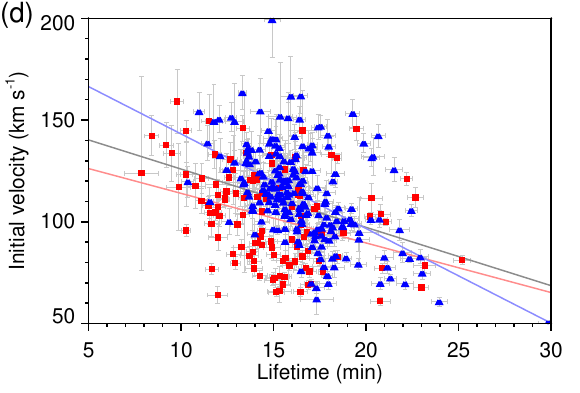}
\hfill
\includegraphics{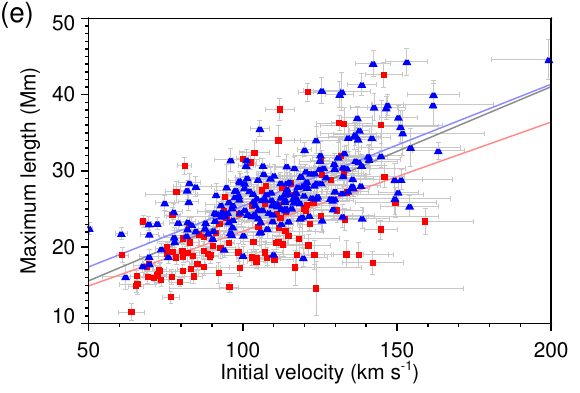}
\hfill
\includegraphics{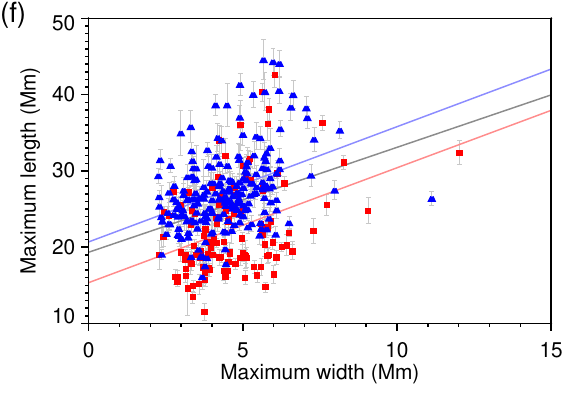}
\\
\includegraphics{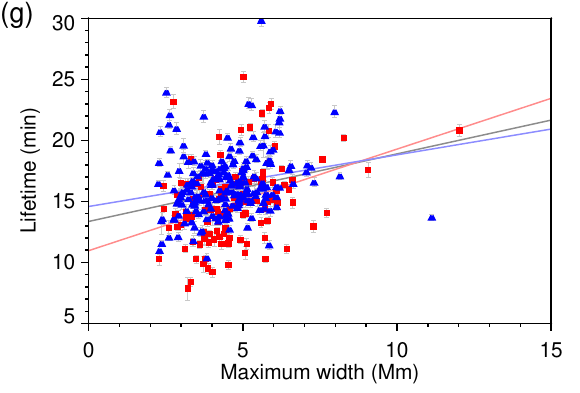}
\hfill
\includegraphics{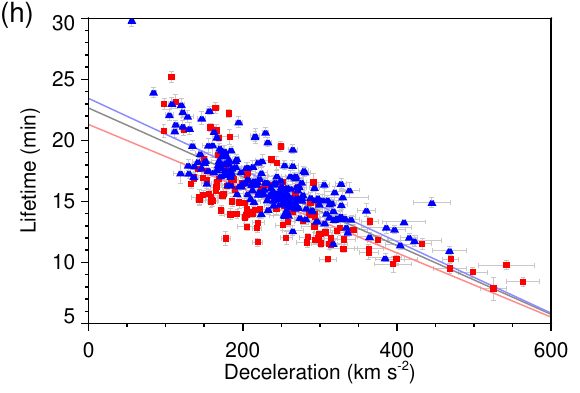}
\hfill
\includegraphics{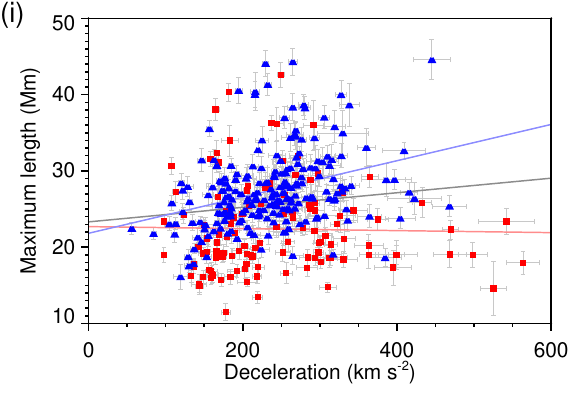}
\caption{Scatter plots for the pairs of jet parameters showing a non-negligible correlation for at least one group of jets. Blue triangles and red squares indicate the coronal hole and quiet-Sun jets respectively. Linear trends for these two groups, and for all of the jets combined, are shown with blue, red, and grey lines respectively. Also note that polar angle in panels (a) and (b) is relative to the nearest pole, and furthermore, is represented in panel (b) by its absolute value.}
\label{fig:corrs}
\end{figure*}

\begin{figure}[t]
\centering
\includegraphics{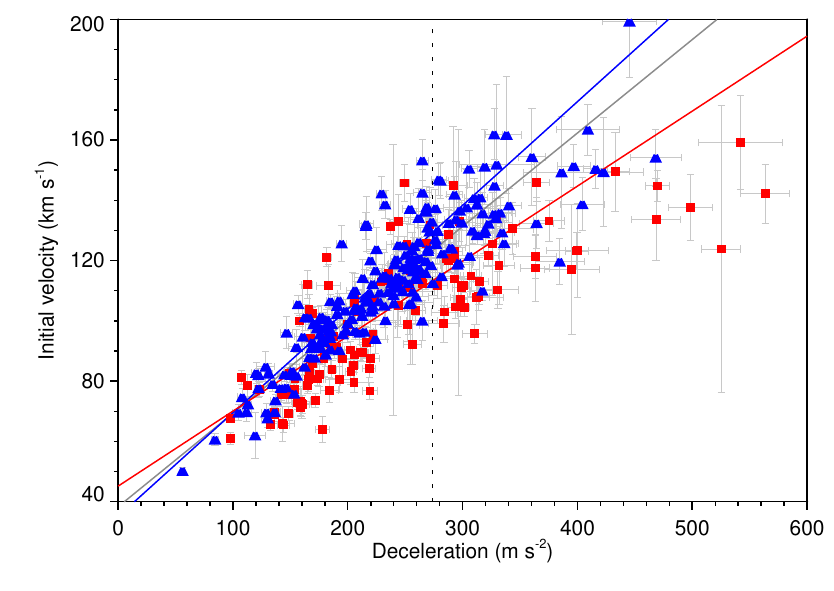}
\caption{Initial velocity of the jets $v_0$ vs. the deceleration $a$. Blue triangles and red squares are for the coronal-hole and quiet-Sun jets, respectively. Linear trends for these two groups, and for all of the jets combined, are shown with blue, red, and grey lines, respectively. The gravitational acceleration at the solar surface $g_{\odot}=247$~m\pss{} is indicated by a vertical dashed line.}
\label{fig:inivel_decel}
\end{figure}

The obtained Pearson's correlation coefficients $r$ for pairs of the main jet characteristics are given in Table~\ref{tbl:corr_par}, while the scatter plots for the most prominent of these correlations are shown in Figs.~\ref{fig:corrs} and \ref{fig:inivel_decel}. 
Therein, if one of the two parameters was strictly positive, while the other could be of either sign, the absolute value of the latter was correspondingly used for the evaluation of their mutual correlation.


To study how the jets' properties are related to their positions on the limb, one needs to set a point of reference for the latter, for which once more we find it useful to define the jet's location relative either to the solar poles or the equator as the possible points of symmetry.
For the coronal-hole jets, we have thus found a moderate ($|r| \geq 0.5$) correlation between the jet's apparent inclination $\alpha$ and the polar angle relative to the nearest pole $\phi_0$ (Fig.~\ref{fig:corrs}a).
For the seemingly linear dependence $\alpha = k \phi_0$, the fitted coefficient is $k=1.01 \pm 0.02 $, which is close to that obtained by \citet{2017A&A...597A..78L} for a smaller set of 36 coronal-hole jets.
This suggests a strong influence of the poloidal global magnetic field in coronal holes on the jets' orientations.
Also, there is a weak ($|r| \geq 0.3$) anticorrelation between the jets' lengths and the distance to the nearest pole (Fig.~\ref{fig:corrs}b), which disappears if each group of jets is considered separately.
This can be certainly explained by the fact that coronal-hole jets, which are typically longer, are concentrated around the solar poles.


Secondly, we observed the naturally expected pairwise correlations between the maximum lengths, lifetimes, and initial velocities of the jets (Figs.~\ref{fig:corrs}c-e).
Indeed, the more energetic jets are likely to reach greater heights and thus spend more time in motion. 
The strengths of the correlations, however, vary for the two groups of jets.
While the jets in the coronal holes show a strong ($|r| \geq 0.7$) correlation between the maximum length and initial velocity, the correlation between the maximum length and lifetime is practically absent.
For the quiet-Sun jets, the strengths of the correlations are, on the contrary, more balanced.
There is also a weak correlation between the length and the width that is equally exhibited by all groups of jets (Fig.~\ref{fig:corrs}f), which shows that longer jets are, after all, generally wider, although not to a great extent.
Finally, for the quiet-Sun jets, there is a weak correlation between the jet's width and its lifetime (Fig.~\ref{fig:corrs}g), which probably arises from the above correlations.


Furthermore, we found no significant correlation between the jet's apparent inclination angle and deceleration.
Although the true inclination of a jet in three dimensions is not known, such a correlation would still have been expected in the ballistic scenario due to the probable symmetry of the jet orientations relative to the radial direction.
Nevertheless, even if the cosine of the inclination angle is considered, the correlation coefficient remains negligibly low at $r=0.04$ (or 0.06 for the coronal-hole and 0.05 for the quiet-Sun jets).
Combined with the fact that the apparent inclinations of the jets are not sufficiently large to explain the lowest decelerations measured, and that a substantial number of the jets have decelerations lager than solar gravity would imply, this clearly indicates that their motion is not purely ballistic.


This context places further emphasis on the found strong anticorrelation of the jet deceleration and lifetime (Fig.~\ref{fig:corrs}i) and an even stronger correlation between the jet deceleration and initial velocity (Fig.~\ref{fig:inivel_decel}).
The latter can be expressed in the form $v_0 = p a + q$, where $p$ and $q$ are the slope and the intercept, respectively, the fitted values for which are given separately for the three groups of jets in Table~\ref{tbl:coeffs}.
Furthermore, both correlations are strongest for the coronal-hole jets.
There is also a weak correlation between the deceleration and the maximum length for the coronal-hole jets (Fig.~\ref{fig:corrs}i), which probably results from the jets' lengths being more strongly correlated with the initial velocities for this group of jets.


Similar correlations have been observed for several types of chromospheric jets: the active-region dynamic fibrils and quiet-Sun mottles on the disk, and the type~I spicules on the limb, which also show parabolic trajectories \citep{2006ApJ...647L..73H, 2007ApJ...655..624D, 2007PASJ...59S.655D, 2007ApJ...660L.169R,2012ApJ...759...18P}.
These authors have shown that such behaviour is well explained by the shock-driven mechanism, which was first put forward by \citet{1982ApJ...257..345H} and \citet{1982SoPh...75...99S}, and could successfully reproduce it in numerical simulations \citep{2007ApJ...666.1277H, 2009ApJ...701.1569M}.
In this model, the jet's plasma follows the N-shaped velocity profile of a slow-mode magneto-acoustic shock, thus exhibiting a linear decrease of velocity, and, correspondingly, the parabolic trajectory of the jet's spire.
In this case, the ``driving force'' discussed at the end of Section~\ref{sect:res_destribs} is ostensibly the increased pressure gradient trailing the shock front.


In the simplest case, the velocity decreases over the shock period $P$ from its initial value $v_0$ to the final $-v_0$ with a constant rate equal to $-a$, which gives rise to a dependence $ 2v_0 = Pa $ \citep{2007ApJ...666.1277H, 2014CEAB...38...39K}.
The presence of a certain dominant period $P_\mathrm{d}$ for a population of jets thus results in a linear correlation between their initial velocity and deceleration, the slope of the dependence being related to this period as $p = P_\mathrm{d}/2$. 
For the jets studied here, this gives a dominant shock period of the order of 10 minutes, with a significant variance over the two kinds of magnetic environment, the period being 39~\% higher for the coronal-hole jets than for the quiet-Sun ones (Table~\ref{tbl:coeffs}).
This dramatically contrasts with the periods measured for the chromospheric jets in the earlier studies, which are at least several times lower (Table~\ref{tbl:chromo}).
This can be partly attributed to the different measurement techniques used: while for the chromospheric jets a maximum observed velocity was considered, in this work the initial velocity was obtained through trajectory extrapolation.
We believe, however, that observational effects are unable to account fully for such a large difference, and a more detailed examination is required.


The above model, however, does not account for the measured non-zero velocity offset $q$, which was also present in observations of chromospheric jets (Tables \ref{tbl:coeffs} and \ref{tbl:chromo}).
This discrepancy is furthermore responsible for the fact that the dominant period measured is significantly lower than the observed mean lifetimes of the jets: otherwise, as follows from the jet's parabolic trajectory, the shock period is $P = 2 v_0/a = t_\mathrm{life}$, \ie{}, exactly the jet's lifetime, which can be seen, for example, from Equations (\ref{eq:lifetime})--(\ref{eq:decel}).
This can be partly explained by the systematic errors in determining the initial velocities (and to a lesser degree, the decelerations) of the jets.
However, we were not able to nullify $q$ by varying the zero height in the velocity extrapolations, thus trying to eliminate the effect of the unknown positions of the jets' footpoints, which we believe exert the strongest influence on our results.
Another possibility is that a modification of the existing theory is required.
For example, a more complex velocity profile of the shock could result in a non-linear dependence in $v_0(a)$, although still producing the nearly parabolic trajectories of the jets' spires.
In this case, however, no unequivocal dominant period can be introduced to explain the existing correlations.

\begin{table}[t]
\centering
\caption{Correlation parameters for $v_0(a)$.} \label{tbl:coeffs}
\begin{tabular}{lM{5.5em}M{5.5em}M{5.5em}}
\hline\hline
Jet location & Slope~$p$ (s) & Intercept~$q$ (km\ps{}) & Period $P_\mathrm{d}$ (min) \\
\hline
All jets		& $310.0 \pm 4.4$ & $38.3 \pm 0.9$ & $10.3 \pm 0.1$ \\
Coronal hole	& $344.0 \pm 5.7$ & $35.1 \pm 1.2$ & $11.5 \pm 0.2$ \\
Quiet Sun		& $248.8 \pm 6.9$ & $45.2 \pm 1.3$ & $8.3 \pm 0.2$ \\
\hline
\end{tabular}
\end{table}

\begin{table}[t]
\caption{Correlation parameters for $v_0(a)$ found in earlier studies of chromospheric jets.} \label{tbl:chromo}
\begin{tabular}{lM{4.5em}M{4.5em}M{4.5em}}
\hline
\hline
Jet type & Slope~$p$\newline(s) & Intercept~$q$\newline(km\ps{}) & Period $P_\mathrm{d}$\newline(min)\\
\hline
Dynamic fibrils\tablenotemark{a,b,c}	&	$60.0 \pm 2.7$ 	&	$9.7 \pm 0.4$ 	&	$2.0 \pm 0.1$ \\
Mottles\tablenotemark{b,c,d}		&	$41.1 \pm 4.5$ 	&	$12.3 \pm 0.7$ 	&	$1.4 \pm 0.1$ \\
Type I spicules\tablenotemark{b,c}	&	$34.0 \pm 7.6$ 	&	$20.9 \pm 2.0$ 	&	$1.1 \pm 0.3$ \\
\hline
\end{tabular}
\tablecomments{The data were extracted from the plots in $^{\mathrm{a}}$\citet{2007ApJ...655..624D}, $^{\mathrm{b}}$\citet{2007PASJ...59S.655D}, $^{\mathrm{c}}$\citet{2012ApJ...759...18P}, and $^{\mathrm{d}}$\citet{2007ApJ...660L.169R}.}
\end{table}

\section{Discussion} \label{sect:discuss}

\subsection{Connection to the chromospheric jets}

In this work, we studied a subset of \euv{} jets that undergo parabolic trajectories in the \ion{He}{2} 304~\AA{} observations.
For each jet, we obtained the main spatial and dynamic characteristics, and studied their distributions and pairwise correlations.
These results have shown that, despite the parabolic trajectories, the dynamics of these jets are inconsistent with a purely ballistic motion.
Moreover, the data suggest that the dynamics of the jets may be governed by magneto-acoustic shocks, which propagate into the corona from the lower layers of the solar atmosphere.
Such a mechanism has been previously proposed to explain similar dynamics in several types of chromospheric jets, including type~I spicules, in which sense the jets studied here are indeed macro-spicules.
However, our results also show a significant difference in the measured dominant periods of the shocks, which suggests a dissimilarity in their formation conditions.
While the type~I spicules were proposed to be triggered by the leakage of the photospheric $p$-mode oscillations into the chromosphere (facilitated by the increased acoustic cut-off period along the inclined magnetic field lines) where they steepen into the magneto-acoustic shocks \citep{1964ApJ...140..288B, 2004Natur.430..536D, 2011ApJ...743..142H, 2011ApJ...743...14S}, this mechanism cannot explain the much longer shock periods observed for the \euv{} jets.

In this respect, it is interesting to note that the more energetic type~II spicules, which undergo vigorous heating and thus fully disappear from the chromospheric spectral lines within 1--2~min, were recently shown to have transition-region counterparts visible in, \eg{}, \ion{He}{2} 304~\AA{} and \ion{Si}{5} 1400~\AA{} lines (both formed at around $8 \times 10^4$~K), which follow nearly parabolic trajectories, thus reaching heights of about 16~Mm and falling back after 8--14 minutes \citep{2014ApJ...792L..15P, 2015ApJ...806..170S}.
It can be therefore expected that these counterparts should be independently identifiable; moreover, the above observations suggest that they should be found among the smallest jets observed in the \euv{}.
Although the reported transition-region counterparts of type II spicules usually have shorter lengths and lifetimes than the average jet studied here, it can be argued that the more energetic jets tend to fade faster from the \ion{Ca}{2}~H line, which was typically used for their identification, and that they therefore could have escaped detection.
Moreover, in these studies, the lifetime was defined by the visibility of the jet, and thus would naturally be smaller than the one obtained by trajectory extrapolation, which additionally covers the unobserved parts of the jet's motion.
Finally, the initial velocities of the jets observed in this study (70--140~km\ps{}) closely match the typical rise velocities reported for the chromospheric type~II spicules, which were found to be in the range of 50--150~km\ps{} \citep{2007PASJ...59S.655D, 2010ApJ...714L...1S, 2012ApJ...752L..12D, 2012ApJ...759...18P}.
Therefore, we suppose at least a partial overlap between the populations of type~II spicules observed in the chromospheric spectral lines and the \euv{} jets studied here; the extent of this overlap, however, requires further examination.

Although the formation mechanism of the type II spicules is still poorly understood, it is generally assumed that this kind of jets is generated by small-scale, low-lying bursts of reconnection, which can provide the sufficiently high initial velocities \citep{2007PASJ...59S.655D, 2010ApJ...714L...1S}.
More recent numerical simulations have also shown that similar jets can be produced by the sporadic release of magnetic tension facilitated by ambipolar diffusion in the partially ionised plasma \citep{2017Sci...356.1269M}.
This model includes the generation of magneto-acoustic shocks, although it is not clear how either this, or the reconnection-driven mechanism, can explain the presence of a certain dominant period for the shocks.
Furthermore, the simulations have shown that type~II spicules can be heated to transition-region, and even to coronal temperatures, and predicted their visibility in the 171~\AA{} and 193~\AA{} channels of the \aia{} \citep{2017ApJ...847...36M, 2018ApJ...860..116M}, which is partly supported by the earlier observations of \citet{2011Sci...331...55D}.
This raises the question of whether the jets studied here have observable coronal counterparts as well.
The observations of macrospicules with parabolic trajectories have shown that, by a rough estimate, from 10 to 30~\% of their cool material fades out from the transition-region spectral line, most likely being heated to higher coronal temperatures, which supports the idea that these jets, and thus presumably the type~II spicules, play a significant role in the mass and energy balance of the corona, and therefore in the formation of the solar wind.

\subsection{Connection to the larger \euv{} jets}

We also note that the jets studied here are among the smallest \euv{} jets observed in the previous works, including those termed macrospicules, which likely results from the inclusion of larger jets due to the different selection criteria.
Interestingly, however, these jets are close to the original macrospicules observed by \citet{1975ApJ...197L.133B}, including their parabolic trajectories.
At the same time, there exists a population of much larger \euv{} jets, such as the transition-region counterparts of blowout coronal jets, some of which which are visible by the C2 instrument of the \lasco{}, thus reaching heights of $1.5 R_\odot$ and above \citep{2008ApJ...680L..73P, 2009SoPh..259...87N,2010AnGeo..28..687N, 2011AdSpR..48.1490N, 2011A&A...534A..62S, 2014A&A...562A..98C, 2015ApJ...806...11M}.
It is not clear, however, whether the difference between these groups of solar jets is merely quantitative or whether they are produced by an essentially different physical mechanism, and in the latter case, whether they can be clearly distinguished by their observed properties, or if there is, for example, any other kind of jets in between.

Observations of larger \euv{} jets, with complex structure and dynamics, have shown that in certain cases, either the jet as a whole or the individual strips and blobs of plasma can also follow parabolic, but often not ballistic, trajectories \citep{2009SoPh..259...87N, 2013SoPh..284..427M, 2014A&A...561A.134Z}.
This implies that the shock-driven mechanism responsible for these parabolic trajectories may not be limited to the relatively simple features studied here, thus being a significantly more widespread phenomenon among solar jets.
Conversely, the precise parabolic trajectories may be observed only in the clearest, unperturbed cases, meaning that the absence of such a trajectory does not necessarily imply a different dynamic mechanism.
Numerical simulations by \citet{2009ApJ...691...61P, 2010ApJ...714.1762P, 2015A&A...573A.130P, 2016A&A...596A..36P} have shown that, depending on how fast the magnetic configuration becomes destabilized, the same set-up can produce either straight, narrow, shock-driven jets sparked by the partial reconnection of the magnetic field lines, or the much larger helical jets resulting from the impulsive break-up and eruption of the entire magnetic structure containing the cool plasma material.
Although aimed primarily at shedding light on the newly found dichotomy of coronal jets, this model also leads us to the idea that the main difference between the two groups of \euv{} jets may be related to whether the associated magnetic configuration remains stable and passively guides the jet's plasma, or erupts with the jet itself, thus giving rise to complex rotary motions.

\subsection{Measurement errors and further study} \label{sect:disc_err}


When it comes to the measurement errors introduced into the above results, one of the major contributing factors is the projection effect, which reduces the observed lengths, inclinations, initial velocities, and decelerations of the jets.
Assuming that jet inclinations along the line of sight are distributed similarly to those in the image plane ($\alpha$), we can evaluate that the above values are on average underestimated by a factor of $\left\langle 1/\cos\vert\alpha\vert \right\rangle = 1.03$, meaning that the error due to the projection effect is about 3~\%.
However, since both the deceleration and initial velocity are equally influenced by this effect, the slope $p$ of the $v_0(a)$ dependence remains unaffected, and so is the inferred dominant period of the shocks, although the measured intercept $q$ is underestimated as well.

The projection effect could possibly be eliminated by viewing the jet from two nearly orthogonal directions, which at the moment can be achieved by stereoscopic observations of the full solar disk offered by the \euvi{} on board the \stereo{}.
However, as the orbits of both \stereo{} spacecrafts are close to the ecliptic plane, such observations would be effective only for the two small regions around the solar poles, where the relatively low-lying jets can be seen off limb from both viewpoints, and moreover would require a reliable cross-identification of these small features.
Additionally, both periods during which the \stereo{} spacecraft were located at an angle of 90\degr{} to each other or to the Earth coincided with the previous minimum of solar activity, and therefore the approximate symmetry of the poloidal magnetic field would likely result in generally predictable inclinations of the polar jets, which are presumably guided by the open magnetic field lines \citep{2017A&A...597A..78L}.

Another way to circumvent the projection effect is to obtain the line-of-sight velocities of the jets through spectroscopic observations of the resulting Doppler shifts.
In  Section~\ref{sect:intro} of this paper, we mentioned some of these measurements performed using the observations of the \cds{} and \sumer{} of the \soho{}; the more up-to-date options also include the \eis{} on board the \textit{Hinode} spacecraft \citep{2007SoPh..243....3K}, the \iris{}, and the planned \textit{EUV Spectroscopic Telescope} (EUVST) of the future \textit{Solar-C} mission \citep{2014SPIE.9143E..1OW, 2016ASPC..504..299S}.
Such instruments, however, typically have a limited temporal and spatial coverage, and thus are poorly suited for extensive statistical studies, being mostly limited to individual case examinations.


The second and probably the most important source of error is rooted in the unknown positions of the jets' footpoints, which are either located behind the limb, or obscured by the optically thick \euv{} spicule ``forest''.
This leaves us to assume that they are located exactly on the limb, and consequently, as the real footpoints are more likely to be located below rather than above the limb, the lengths of the jets are possibly underestimated, as well as their initial velocities and lifetimes, which are obtained through the trajectory extrapolation.
This, however, does not affect the observed decelerations, meaning that the slope of the $v_0(a)$ dependence should be somewhat steeper, thus making the difference of the shock periods driving the \euv{} and chromospheric jets even more pronounced.

To give a numerical estimate of this effect, we performed a Monte Carlo simulation using a model population of jets, which were scattered evenly across the surface of the Sun and had a Gaussian distribution of their lengths.
Considering only those jets whose lengths allow them to be effectively observed above the EUV spicule "forest", the distribution of the apparent lengths above the limb was calculated, after which  parameters of the model were tuned to match the observed distribution.
We have thus inferred that, on average, the lengths of the jets are underestimated by around 14~\%, which makes this effect the main contributing factor to the error of our results.
Following Equations (\ref{eq:hmax})--(\ref{eq:inivel}), the corresponding errors of the jets' lifetimes and initial velocities can be evaluated as around 7~\%.


Finally, the non-uniform observational coverage of the dataset does not allow us to consistently track the possible variations of the jets' properties throughout the solar cycle. 
We therefore plan to further employ the vast pool of \aia{} data to perform a more extensive study of the different kinds of \euv{} jets by combining a well-organized set of multi-wavelength observations in the transition-region and coronal spectral lines with the data reduction method described in Section~\ref{sect:data_methods}, which enables efficient image processing with a minimum of manual operations.
This approach will hopefully shed more light on the interrelation of different types of jets, thus helping to establish better-defined boundaries between their sub-populations based on the different underlying physical mechanisms rather than on their apparent size and morphology.
Furthermore, such a study will enable us to detect the presence of hot plasma components for each type of jets, and thus to evaluate their relative impact on the mass and energy balance of the corona.

Future studies should also address the identification of the on-disk counterparts of the jets described here by taking advantage of modern-era high-resolution and high-cadence observations.
The on-disk observations of the jets will eventually bring a new understanding of jet formation mechanisms, thanks to the availability of photospheric magnetic field measurements such as those provided by the \hmi{} of the \sdo{}, which should give further insight into the magnetic configuration of the jets and the possible magnetic drivers.
Furthermore, an extensive cross-correlation study is needed to evaluate the extent of the assumed overlap between the populations of type II spicules and \euv{} jets with parabolic trajectories, involving the independent identification of these features in both transition-region and chromospheric spectral lines.
Finally, the origin of the dominant period of the assumed magneto-acoustic shocks should be convincingly explained, either aided by new observations or with the help of numerical simulations, which should be able to reproduce the linear character of the $v_0(a)$ dependence observed in this study.

\acknowledgments
This work was supported by the Russian Science Foundation (grant 17-12-01567).
The authors would like to thank all the contributors to the \aia{} branch of the IDL \textit{SolarSoft} package and equally those who have developed this outstanding instrument and maintain its operation on a daily basis.
I.P.L. is also grateful to A.~C. Stirling for providing the headline question, and thus motivating a significant portion of this research.

\bibliography{WiaM}

\begin{thebibliography}{}
\expandafter\ifx\csname natexlab\endcsname\relax\def\natexlab#1{#1}\fi
\providecommand{\url}[1]{\href{#1}{#1}}
\providecommand{\dodoi}[1]{doi:~\href{http://doi.org/#1}{\nolinkurl{#1}}}
\providecommand{\doeprint}[1]{\href{http://ascl.net/#1}{\nolinkurl{http://ascl.net/#1}}}
\providecommand{\doarXiv}[1]{\href{https://arxiv.org/abs/#1}{\nolinkurl{https://arxiv.org/abs/#1}}}

\bibitem[{{Adams} {et~al.}(2014){Adams}, {Sterling}, {Moore}, \&
  {Gary}}]{2014ApJ...783...11A}
{Adams}, M., {Sterling}, A.~C., {Moore}, R.~L., \& {Gary}, G.~A. 2014, \apj,
  783, 11, \dodoi{10.1088/0004-637X/783/1/11}

\bibitem[{{Andreev} \& {Kosovichev}(1994)}]{1994ESASP.373..179A}
{Andreev}, A.~S., \& {Kosovichev}, A.~G. 1994, in ESA Special Publication, Vol.
  373, Solar Dynamic Phenomena and Solar Wind Consequences, the Third SOHO
  Workshop, ed. J.~J. {Hunt}, 179--182

\bibitem[{{Banerjee} {et~al.}(2000){Banerjee}, {O'Shea}, \&
  {Doyle}}]{2000A&A...355.1152B}
{Banerjee}, D., {O'Shea}, E., \& {Doyle}, J.~G. 2000, \aap, 355, 1152

\bibitem[{{Beckers}(1968)}]{1968SoPh....3..367B}
{Beckers}, J.~M. 1968, \solphys, 3, 367, \dodoi{10.1007/BF00171614}

\bibitem[{{Beckers}(1972)}]{1972ARA&A..10...73B}
---. 1972, \araa, 10, 73, \dodoi{10.1146/annurev.aa.10.090172.000445}

\bibitem[{{Bennett} \& {Erd{\'e}lyi}(2015)}]{2015ApJ...808..135B}
{Bennett}, S.~M., \& {Erd{\'e}lyi}, R. 2015, \apj, 808, 135,
  \dodoi{10.1088/0004-637X/808/2/135}

\bibitem[{{Bird}(1964)}]{1964ApJ...140..288B}
{Bird}, G.~A. 1964, \apj, 140, 288, \dodoi{10.1086/147917}

\bibitem[{{Blake} \& {Sturrock}(1985)}]{1985ApJ...290..359B}
{Blake}, M.~L., \& {Sturrock}, P.~A. 1985, \apj, 290, 359,
  \dodoi{10.1086/162993}

\bibitem[{{Boerner} {et~al.}(2014){Boerner}, {Testa}, {Warren}, {Weber}, \&
  {Schrijver}}]{2014SoPh..289.2377B}
{Boerner}, P.~F., {Testa}, P., {Warren}, H., {Weber}, M.~A., \& {Schrijver},
  C.~J. 2014, \solphys, 289, 2377, \dodoi{10.1007/s11207-013-0452-z}

\bibitem[{{Bohlin} {et~al.}(1975){Bohlin}, {Vogel}, {Purcell}, {Sheeley},
  {Tousey}, \& {Vanhoosier}}]{1975ApJ...197L.133B}
{Bohlin}, J.~D., {Vogel}, S.~N., {Purcell}, J.~D., {et~al.} 1975, \apjl, 197,
  L133, \dodoi{10.1086/181794}

\bibitem[{{Brueckner} {et~al.}(1995){Brueckner}, {Howard}, {Koomen},
  {Korendyke}, {Michels}, {Moses}, {Socker}, {Dere}, {Lamy}, {Llebaria},
  {Bout}, {Schwenn}, {Simnett}, {Bedford}, \& {Eyles}}]{1995SoPh..162..357B}
{Brueckner}, G.~E., {Howard}, R.~A., {Koomen}, M.~J., {et~al.} 1995, \solphys,
  162, 357, \dodoi{10.1007/BF00733434}

\bibitem[{{Budnik} {et~al.}(1998){Budnik}, {Schroeder}, {Wilhelm}, \&
  {Glassmeier}}]{1998A&A...334L..77B}
{Budnik}, F., {Schroeder}, K.-P., {Wilhelm}, K., \& {Glassmeier}, K.-H. 1998,
  \aap, 334, L77

\bibitem[{{Chandrashekhar} {et~al.}(2014){Chandrashekhar}, {Morton},
  {Banerjee}, \& {Gupta}}]{2014A&A...562A..98C}
{Chandrashekhar}, K., {Morton}, R.~J., {Banerjee}, D., \& {Gupta}, G.~R. 2014,
  \aap, 562, A98, \dodoi{10.1051/0004-6361/201322408}

\bibitem[{{Culhane} {et~al.}(2007){Culhane}, {Harra}, {James}, {Al-Janabi},
  {Bradley}, {Chaudry}, {Rees}, {Tandy}, {Thomas}, {Whillock}, {Winter},
  {Doschek}, {Korendyke}, {Brown}, {Myers}, {Mariska}, {Seely}, {Lang}, {Kent},
  {Shaughnessy}, {Young}, {Simnett}, {Castelli}, {Mahmoud}, {Mapson-Menard},
  {Probyn}, {Thomas}, {Davila}, {Dere}, {Windt}, {Shea}, {Hagood}, {Moye},
  {Hara}, {Watanabe}, {Matsuzaki}, {Kosugi}, {Hansteen}, \&
  {Wikstol}}]{2007SoPh..243...19C}
{Culhane}, J.~L., {Harra}, L.~K., {James}, A.~M., {et~al.} 2007, \solphys, 243,
  19, \dodoi{10.1007/s01007-007-0293-1}

\bibitem[{{Cushman} \& {Rense}(1978)}]{1978SoPh...58..299C}
{Cushman}, G.~W., \& {Rense}, W.~A. 1978, \solphys, 58, 299,
  \dodoi{10.1007/BF00157275}

\bibitem[{{de Pontieu} {et~al.}(2012){de Pontieu}, {Carlsson}, {Rouppe van der
  Voort}, {Rutten}, {Hansteen}, \& {Watanabe}}]{2012ApJ...752L..12D}
{de Pontieu}, B., {Carlsson}, M., {Rouppe van der Voort}, L.~H.~M., {et~al.}
  2012, \apjl, 752, L12, \dodoi{10.1088/2041-8205/752/1/L12}

\bibitem[{{de Pontieu} {et~al.}(2004){de Pontieu}, {Erd{\'e}lyi}, \&
  {James}}]{2004Natur.430..536D}
{de Pontieu}, B., {Erd{\'e}lyi}, R., \& {James}, S.~P. 2004, \nat, 430, 536,
  \dodoi{10.1038/nature02749}

\bibitem[{{de Pontieu} {et~al.}(2007{\natexlab{a}}){de Pontieu}, {Hansteen},
  {Rouppe van der Voort}, {van Noort}, \& {Carlsson}}]{2007ApJ...655..624D}
{de Pontieu}, B., {Hansteen}, V.~H., {Rouppe van der Voort}, L., {van Noort},
  M., \& {Carlsson}, M. 2007{\natexlab{a}}, \apj, 655, 624,
  \dodoi{10.1086/509070}

\bibitem[{{de Pontieu} {et~al.}(2007{\natexlab{b}}){de Pontieu}, {McIntosh},
  {Hansteen}, {Carlsson}, {Schrijver}, {Tarbell}, {Title}, {Shine}, {Suematsu},
  {Tsuneta}, {Katsukawa}, {Ichimoto}, {Shimizu}, \&
  {Nagata}}]{2007PASJ...59S.655D}
{de Pontieu}, B., {McIntosh}, S., {Hansteen}, V.~H., {et~al.}
  2007{\natexlab{b}}, \pasj, 59, S655.
\newblock \doarXiv{0710.2934}

\bibitem[{{de Pontieu} {et~al.}(2011){de Pontieu}, {McIntosh}, {Carlsson},
  {Hansteen}, {Tarbell}, {Boerner}, {Martinez-Sykora}, {Schrijver}, \&
  {Title}}]{2011Sci...331...55D}
{de Pontieu}, B., {McIntosh}, S.~W., {Carlsson}, M., {et~al.} 2011, Science,
  331, 55, \dodoi{10.1126/science.1197738}

\bibitem[{{de Pontieu} {et~al.}(2014){de Pontieu}, {Title}, {Lemen}, {Kushner},
  {Akin}, {Allard}, {Berger}, {Boerner}, {Cheung}, {Chou}, {Drake}, {Duncan},
  {Freeland}, {Heyman}, {Hoffman}, {Hurlburt}, {Lindgren}, {Mathur}, {Rehse},
  {Sabolish}, {Seguin}, {Schrijver}, {Tarbell}, {W{\"u}lser}, {Wolfson},
  {Yanari}, {Mudge}, {Nguyen-Phuc}, {Timmons}, {van Bezooijen}, {Weingrod},
  {Brookner}, {Butcher}, {Dougherty}, {Eder}, {Knagenhjelm}, {Larsen},
  {Mansir}, {Phan}, {Boyle}, {Cheimets}, {DeLuca}, {Golub}, {Gates}, {Hertz},
  {McKillop}, {Park}, {Perry}, {Podgorski}, {Reeves}, {Saar}, {Testa}, {Tian},
  {Weber}, {Dunn}, {Eccles}, {Jaeggli}, {Kankelborg}, {Mashburn}, {Pust},
  {Springer}, {Carvalho}, {Kleint}, {Marmie}, {Mazmanian}, {Pereira}, {Sawyer},
  {Strong}, {Worden}, {Carlsson}, {Hansteen}, {Leenaarts}, {Wiesmann},
  {Aloise}, {Chu}, {Bush}, {Scherrer}, {Brekke}, {Martinez-Sykora}, {Lites},
  {McIntosh}, {Uitenbroek}, {Okamoto}, {Gummin}, {Auker}, {Jerram}, {Pool}, \&
  {Waltham}}]{2014SoPh..289.2733D}
{de Pontieu}, B., {Title}, A.~M., {Lemen}, J.~R., {et~al.} 2014, \solphys, 289,
  2733, \dodoi{10.1007/s11207-014-0485-y}

\bibitem[{{Delaboudini{\`e}re}(1999)}]{1999SoPh..188..259D}
{Delaboudini{\`e}re}, J.~P. 1999, \solphys, 188, 259,
  \dodoi{10.1023/A:1005268807716}

\bibitem[{{Dere} {et~al.}(1983){Dere}, {Bartoe}, \&
  {Brueckner}}]{1983ApJ...267L..65D}
{Dere}, K.~P., {Bartoe}, J.-D.~F., \& {Brueckner}, G.~E. 1983, \apjl, 267, L65,
  \dodoi{10.1086/184004}

\bibitem[{{Dere} {et~al.}(1989){Dere}, {Bartoe}, {Brueckner}, {Cook}, \&
  {Socker}}]{1989SoPh..119...55D}
{Dere}, K.~P., {Bartoe}, J.-D.~F., {Brueckner}, G.~E., {Cook}, J.~W., \&
  {Socker}, D.~G. 1989, \solphys, 119, 55, \dodoi{10.1007/BF00146212}

\bibitem[{{Domingo} {et~al.}(1995){Domingo}, {Fleck}, \&
  {Poland}}]{1995SoPh..162....1D}
{Domingo}, V., {Fleck}, B., \& {Poland}, A.~I. 1995, \solphys, 162, 1,
  \dodoi{10.1007/BF00733425}

\bibitem[{{Georgakilas} {et~al.}(1999){Georgakilas}, {Koutchmy}, \&
  {Alissandrakis}}]{1999A&A...341..610G}
{Georgakilas}, A.~A., {Koutchmy}, S., \& {Alissandrakis}, C.~E. 1999, \aap,
  341, 610

\bibitem[{{Godoli} \& {Mazzucconi}(1967)}]{1967ApJ...147.1131G}
{Godoli}, G., \& {Mazzucconi}, F. 1967, \apj, 147, 1131, \dodoi{10.1086/149102}

\bibitem[{{Gyenge} {et~al.}(2015){Gyenge}, {Bennett}, \&
  {Erd{\'e}lyi}}]{2015JApA...36..103G}
{Gyenge}, N., {Bennett}, S., \& {Erd{\'e}lyi}, R. 2015, Journal of Astrophysics
  and Astronomy, 36, 103, \dodoi{10.1007/s12036-015-9316-2}

\bibitem[{{Habbal} \& {Gonzalez}(1991)}]{1991ApJ...376L..25H}
{Habbal}, S.~R., \& {Gonzalez}, R.~D. 1991, \apjl, 376, L25,
  \dodoi{10.1086/186094}

\bibitem[{{Hansteen} {et~al.}(2006){Hansteen}, {De Pontieu}, {Rouppe van der
  Voort}, {van Noort}, \& {Carlsson}}]{2006ApJ...647L..73H}
{Hansteen}, V.~H., {De Pontieu}, B., {Rouppe van der Voort}, L., {van Noort},
  M., \& {Carlsson}, M. 2006, \apjl, 647, L73, \dodoi{10.1086/507452}

\bibitem[{{Harrison}(1997)}]{1997SoPh..175..467H}
{Harrison}, R.~A. 1997, \solphys, 175, 467, \dodoi{10.1023/A:1004964707047}

\bibitem[{{Harrison} {et~al.}(1995){Harrison}, {Sawyer}, {Carter}, {Cruise},
  {Cutler}, {Fludra}, {Hayes}, {Kent}, {Lang}, {Parker}, {Payne}, {Pike},
  {Peskett}, {Richards}, {Gulhane}, {Norman}, {Breeveld}, {Breeveld}, {Al
  Janabi}, {McCalden}, {Parkinson}, {Self}, {Thomas}, {Poland}, {Thomas},
  {Thompson}, {Kjeldseth-Moe}, {Brekke}, {Karud}, {Maltby}, {Aschenbach},
  {Br{\"a}uninger}, {K{\"u}hne}, {Hollandt}, {Siegmund}, {Huber}, {Gabriel},
  {Mason}, \& {Bromage}}]{1995SoPh..162..233H}
{Harrison}, R.~A., {Sawyer}, E.~C., {Carter}, M.~K., {et~al.} 1995, \solphys,
  162, 233, \dodoi{10.1007/BF00733431}

\bibitem[{{Heggland} {et~al.}(2007){Heggland}, {De Pontieu}, \&
  {Hansteen}}]{2007ApJ...666.1277H}
{Heggland}, L., {De Pontieu}, B., \& {Hansteen}, V.~H. 2007, \apj, 666, 1277,
  \dodoi{10.1086/518828}

\bibitem[{{Heggland} {et~al.}(2011){Heggland}, {Hansteen}, {De Pontieu}, \&
  {Carlsson}}]{2011ApJ...743..142H}
{Heggland}, L., {Hansteen}, V.~H., {De Pontieu}, B., \& {Carlsson}, M. 2011,
  \apj, 743, 142, \dodoi{10.1088/0004-637X/743/2/142}

\bibitem[{{Hermans} \& {Martin}(1986)}]{1986NASCP2442..369H}
{Hermans}, L.~M., \& {Martin}, S.~F. 1986, in NASA Conference Publication, Vol.
  2442, NASA Conference Publication, ed. A.~I. {Poland}

\bibitem[{{Hollweg}(1982)}]{1982ApJ...257..345H}
{Hollweg}, J.~V. 1982, \apj, 257, 345, \dodoi{10.1086/159993}

\bibitem[{{Jelinsky} {et~al.}(1995){Jelinsky}, {Vallerga}, \&
  {Edelstein}}]{1995ApJ...442..653J}
{Jelinsky}, P., {Vallerga}, J.~V., \& {Edelstein}, J. 1995, \apj, 442, 653,
  \dodoi{10.1086/175469}

\bibitem[{{Jordan}(1975)}]{1975MNRAS.170..429J}
{Jordan}, C. 1975, \mnras, 170, 429, \dodoi{10.1093/mnras/170.2.429}

\bibitem[{{Kaiser} {et~al.}(2008){Kaiser}, {Kucera}, {Davila}, {St.~Cyr},
  {Guhathakurta}, \& {Christian}}]{2008SSRv..136....5K}
{Kaiser}, M.~L., {Kucera}, T.~A., {Davila}, J.~M., {et~al.} 2008, \ssr, 136, 5,
  \dodoi{10.1007/s11214-007-9277-0}

\bibitem[{{Kamio} {et~al.}(2010){Kamio}, {Curdt}, {Teriaca}, {Inhester}, \&
  {Solanki}}]{2010A&A...510L...1K}
{Kamio}, S., {Curdt}, W., {Teriaca}, L., {Inhester}, B., \& {Solanki}, S.~K.
  2010, \aap, 510, L1, \dodoi{10.1051/0004-6361/200913269}

\bibitem[{{Karovska} \& {Habbal}(1994)}]{1994ApJ...431L..59K}
{Karovska}, M., \& {Habbal}, S.~R. 1994, \apjl, 431, L59,
  \dodoi{10.1086/187472}

\bibitem[{{Kayshap} {et~al.}(2013){Kayshap}, {Srivastava}, {Murawski}, \&
  {Tripathi}}]{2013ApJ...770L...3K}
{Kayshap}, P., {Srivastava}, A.~K., {Murawski}, K., \& {Tripathi}, D. 2013,
  \apjl, 770, L3, \dodoi{10.1088/2041-8205/770/1/L3}

\bibitem[{Keys(1981)}]{keys1981cubic}
Keys, R. 1981, IEEE transactions on acoustics, speech, and signal processing,
  29, 1153, \dodoi{10.1109/TASSP.1981.1163711}

\bibitem[{{Kiss} \& {Erd{\'e}lyi}(2018)}]{2018ApJ...857..113K}
{Kiss}, T.~S., \& {Erd{\'e}lyi}, R. 2018, \apj, 857, 113,
  \dodoi{10.3847/1538-4357/aab8f7}

\bibitem[{{Kiss} {et~al.}(2017){Kiss}, {Gyenge}, \&
  {Erd{\'e}lyi}}]{2017ApJ...835...47K}
{Kiss}, T.~S., {Gyenge}, N., \& {Erd{\'e}lyi}, R. 2017, \apj, 835, 47,
  \dodoi{10.3847/1538-4357/aa5272}

\bibitem[{{Kiss} {et~al.}(2018){Kiss}, {Gyenge}, \&
  {Erd{\'e}lyi}}]{2018AdSpR..61..611K}
---. 2018, Advances in Space Research, 61, 611,
  \dodoi{10.1016/j.asr.2017.05.027}

\bibitem[{{Kjeldseth-Moe} {et~al.}(1975){Kjeldseth-Moe}, {Beckers}, \&
  {Engvold}}]{1975SoPh...40...65M}
{Kjeldseth-Moe}, O., {Beckers}, J.~M., \& {Engvold}, O. 1975, \solphys, 40, 65,
  \dodoi{10.1007/BF00183152}

\bibitem[{{Kosugi} {et~al.}(2007){Kosugi}, {Matsuzaki}, {Sakao}, {Shimizu},
  {Sone}, {Tachikawa}, {Hashimoto}, {Minesugi}, {Ohnishi}, {Yamada}, {Tsuneta},
  {Hara}, {Ichimoto}, {Suematsu}, {Shimojo}, {Watanabe}, {Shimada}, {Davis},
  {Hill}, {Owens}, {Title}, {Culhane}, {Harra}, {Doschek}, \&
  {Golub}}]{2007SoPh..243....3K}
{Kosugi}, T., {Matsuzaki}, K., {Sakao}, T., {et~al.} 2007, \solphys, 243, 3,
  \dodoi{10.1007/s11207-007-9014-6}

\bibitem[{{Koza}(2014)}]{2014CEAB...38...39K}
{Koza}, J. 2014, Central European Astrophysical Bulletin, 38, 39

\bibitem[{{LaBonte}(1979)}]{1979SoPh...61..283L}
{LaBonte}, B.~J. 1979, \solphys, 61, 283, \dodoi{10.1007/BF00150413}

\bibitem[{{Labrosse} \& {Gouttebroze}(2001)}]{2001A&A...380..323L}
{Labrosse}, N., \& {Gouttebroze}, P. 2001, \aap, 380, 323,
  \dodoi{10.1051/0004-6361:20011395}

\bibitem[{{Lemen} {et~al.}(2012){Lemen}, {Title}, {Akin}, {Boerner}, {Chou},
  {Drake}, {Duncan}, {Edwards}, {Friedlaender}, {Heyman}, {Hurlburt}, {Katz},
  {Kushner}, {Levay}, {Lindgren}, {Mathur}, {McFeaters}, {Mitchell}, {Rehse},
  {Schrijver}, {Springer}, {Stern}, {Tarbell}, {Wuelser}, {Wolfson}, {Yanari},
  {Bookbinder}, {Cheimets}, {Caldwell}, {Deluca}, {Gates}, {Golub}, {Park},
  {Podgorski}, {Bush}, {Scherrer}, {Gummin}, {Smith}, {Auker}, {Jerram},
  {Pool}, {Soufli}, {Windt}, {Beardsley}, {Clapp}, {Lang}, \&
  {Waltham}}]{2012SoPh..275...17L}
{Lemen}, J.~R., {Title}, A.~M., {Akin}, D.~J., {et~al.} 2012, \solphys, 275,
  17, \dodoi{10.1007/s11207-011-9776-8}

\bibitem[{{Loboda} \& {Bogachev}(2015)}]{2015SoPh..290.1963L}
{Loboda}, I.~P., \& {Bogachev}, S.~A. 2015, \solphys, 290, 1963,
  \dodoi{10.1007/s11207-015-0735-7}

\bibitem[{{Loboda} \& {Bogachev}(2017)}]{2017A&A...597A..78L}
---. 2017, \aap, 597, A78, \dodoi{10.1051/0004-6361/201527559}

\bibitem[{{Loucif}(1994)}]{1994A&A...281...95L}
{Loucif}, M.~L. 1994, \aap, 281, 95

\bibitem[{{Madjarska} {et~al.}(2006){Madjarska}, {Doyle}, {Hochedez}, \&
  {Theissen}}]{2006A&A...452L..11M}
{Madjarska}, M.~S., {Doyle}, J.~G., {Hochedez}, J.-F., \& {Theissen}, A. 2006,
  \aap, 452, L11, \dodoi{10.1051/0004-6361:200600024}

\bibitem[{{Madjarska} {et~al.}(2011){Madjarska}, {Vanninathan}, \&
  {Doyle}}]{2011A&A...532L...1M}
{Madjarska}, M.~S., {Vanninathan}, K., \& {Doyle}, J.~G. 2011, \aap, 532, L1,
  \dodoi{10.1051/0004-6361/201116735}

\bibitem[{{Mart{\'{\i}}nez-Sykora}
  {et~al.}(2017{\natexlab{a}}){Mart{\'{\i}}nez-Sykora}, {De Pontieu},
  {Carlsson}, {Hansteen}, {N{\'o}brega-Siverio}, \&
  {Gudiksen}}]{2017ApJ...847...36M}
{Mart{\'{\i}}nez-Sykora}, J., {De Pontieu}, B., {Carlsson}, M., {et~al.}
  2017{\natexlab{a}}, \apj, 847, 36, \dodoi{10.3847/1538-4357/aa8866}

\bibitem[{{Mart{\'{\i}}nez-Sykora} {et~al.}(2018){Mart{\'{\i}}nez-Sykora}, {De
  Pontieu}, {De Moortel}, {Hansteen}, \& {Carlsson}}]{2018ApJ...860..116M}
{Mart{\'{\i}}nez-Sykora}, J., {De Pontieu}, B., {De Moortel}, I., {Hansteen},
  V.~H., \& {Carlsson}, M. 2018, \apj, 860, 116,
  \dodoi{10.3847/1538-4357/aac2ca}

\bibitem[{{Mart{\'{\i}}nez-Sykora}
  {et~al.}(2017{\natexlab{b}}){Mart{\'{\i}}nez-Sykora}, {De Pontieu},
  {Hansteen}, {Rouppe van der Voort}, {Carlsson}, \&
  {Pereira}}]{2017Sci...356.1269M}
{Mart{\'{\i}}nez-Sykora}, J., {De Pontieu}, B., {Hansteen}, V.~H., {et~al.}
  2017{\natexlab{b}}, Science, 356, 1269, \dodoi{10.1126/science.aah5412}

\bibitem[{{Mart{\'{\i}}nez-Sykora} {et~al.}(2009){Mart{\'{\i}}nez-Sykora},
  {Hansteen}, {De Pontieu}, \& {Carlsson}}]{2009ApJ...701.1569M}
{Mart{\'{\i}}nez-Sykora}, J., {Hansteen}, V., {De Pontieu}, B., \& {Carlsson},
  M. 2009, \apj, 701, 1569, \dodoi{10.1088/0004-637X/701/2/1569}

\bibitem[{{Moore} {et~al.}(2010){Moore}, {Cirtain}, {Sterling}, \&
  {Falconer}}]{2010ApJ...720..757M}
{Moore}, R.~L., {Cirtain}, J.~W., {Sterling}, A.~C., \& {Falconer}, D.~A. 2010,
  \apj, 720, 757, \dodoi{10.1088/0004-637X/720/1/757}

\bibitem[{{Moore} {et~al.}(2015){Moore}, {Sterling}, \&
  {Falconer}}]{2015ApJ...806...11M}
{Moore}, R.~L., {Sterling}, A.~C., \& {Falconer}, D.~A. 2015, \apj, 806, 11,
  \dodoi{10.1088/0004-637X/806/1/11}

\bibitem[{{Moore} {et~al.}(2013){Moore}, {Sterling}, {Falconer}, \&
  {Robe}}]{2013ApJ...769..134M}
{Moore}, R.~L., {Sterling}, A.~C., {Falconer}, D.~A., \& {Robe}, D. 2013, \apj,
  769, 134, \dodoi{10.1088/0004-637X/769/2/134}

\bibitem[{{Moore} {et~al.}(1977){Moore}, {Tang}, {Bohlin}, \&
  {Golub}}]{1977ApJ...218..286M}
{Moore}, R.~L., {Tang}, F., {Bohlin}, J.~D., \& {Golub}, L. 1977, \apj, 218,
  286, \dodoi{10.1086/155681}

\bibitem[{{Moschou} {et~al.}(2013){Moschou}, {Tsinganos}, {Vourlidas}, \&
  {Archontis}}]{2013SoPh..284..427M}
{Moschou}, S.~P., {Tsinganos}, K., {Vourlidas}, A., \& {Archontis}, V. 2013,
  \solphys, 284, 427, \dodoi{10.1007/s11207-012-0190-7}

\bibitem[{{Murawski} {et~al.}(2011){Murawski}, {Srivastava}, \&
  {Zaqarashvili}}]{2011A&A...535A..58M}
{Murawski}, K., {Srivastava}, A.~K., \& {Zaqarashvili}, T.~V. 2011, \aap, 535,
  A58, \dodoi{10.1051/0004-6361/201117589}

\bibitem[{{Nistic{\`o}} {et~al.}(2009){Nistic{\`o}}, {Bothmer}, {Patsourakos},
  \& {Zimbardo}}]{2009SoPh..259...87N}
{Nistic{\`o}}, G., {Bothmer}, V., {Patsourakos}, S., \& {Zimbardo}, G. 2009,
  \solphys, 259, 87, \dodoi{10.1007/s11207-009-9424-8}

\bibitem[{{Nistic{\`o}} {et~al.}(2010){Nistic{\`o}}, {Bothmer}, {Patsourakos},
  \& {Zimbardo}}]{2010AnGeo..28..687N}
---. 2010, Annales Geophysicae, 28, 687, \dodoi{10.5194/angeo-28-687-2010}

\bibitem[{{Nistic{\`o}} {et~al.}(2011){Nistic{\`o}}, {Patsourakos}, {Bothmer},
  \& {Zimbardo}}]{2011AdSpR..48.1490N}
{Nistic{\`o}}, G., {Patsourakos}, S., {Bothmer}, V., \& {Zimbardo}, G. 2011,
  Advances in Space Research, 48, 1490, \dodoi{10.1016/j.asr.2011.07.003}

\bibitem[{{O'Shea} {et~al.}(2005){O'Shea}, {Banerjee}, \&
  {Doyle}}]{2005A&A...436L..43O}
{O'Shea}, E., {Banerjee}, D., \& {Doyle}, J.~G. 2005, \aap, 436, L43,
  \dodoi{10.1051/0004-6361:200500119}

\bibitem[{{Parenti} {et~al.}(2002){Parenti}, {Bromage}, \&
  {Bromage}}]{2002A&A...384..303P}
{Parenti}, S., {Bromage}, B.~J.~I., \& {Bromage}, G.~E. 2002, \aap, 384, 303,
  \dodoi{10.1051/0004-6361:20011819}

\bibitem[{{Pariat} {et~al.}(2009){Pariat}, {Antiochos}, \&
  {DeVore}}]{2009ApJ...691...61P}
{Pariat}, E., {Antiochos}, S.~K., \& {DeVore}, C.~R. 2009, \apj, 691, 61,
  \dodoi{10.1088/0004-637X/691/1/61}

\bibitem[{{Pariat} {et~al.}(2010){Pariat}, {Antiochos}, \&
  {DeVore}}]{2010ApJ...714.1762P}
---. 2010, \apj, 714, 1762, \dodoi{10.1088/0004-637X/714/2/1762}

\bibitem[{{Pariat} {et~al.}(2015){Pariat}, {Dalmasse}, {DeVore}, {Antiochos},
  \& {Karpen}}]{2015A&A...573A.130P}
{Pariat}, E., {Dalmasse}, K., {DeVore}, C.~R., {Antiochos}, S.~K., \& {Karpen},
  J.~T. 2015, \aap, 573, A130, \dodoi{10.1051/0004-6361/201424209}

\bibitem[{{Pariat} {et~al.}(2016){Pariat}, {Dalmasse}, {DeVore}, {Antiochos},
  \& {Karpen}}]{2016A&A...596A..36P}
---. 2016, \aap, 596, A36, \dodoi{10.1051/0004-6361/201629109}

\bibitem[{Park \& Schowengerdt(1983)}]{PARK1983258}
Park, S.~K., \& Schowengerdt, R.~A. 1983, Computer Vision, Graphics, and Image
  Processing, 23, 258 , \dodoi{http://dx.doi.org/10.1016/0734-189X(83)90026-9}

\bibitem[{{Pasachoff} {et~al.}(2009){Pasachoff}, {Jacobson}, \&
  {Sterling}}]{2009SoPh..260...59P}
{Pasachoff}, J.~M., {Jacobson}, W.~A., \& {Sterling}, A.~C. 2009, \solphys,
  260, 59, \dodoi{10.1007/s11207-009-9430-x}

\bibitem[{{Patsourakos} {et~al.}(2008){Patsourakos}, {Pariat}, {Vourlidas},
  {Antiochos}, \& {Wuelser}}]{2008ApJ...680L..73P}
{Patsourakos}, S., {Pariat}, E., {Vourlidas}, A., {Antiochos}, S.~K., \&
  {Wuelser}, J.~P. 2008, \apjl, 680, L73, \dodoi{10.1086/589769}

\bibitem[{{Pereira} {et~al.}(2012){Pereira}, {De Pontieu}, \&
  {Carlsson}}]{2012ApJ...759...18P}
{Pereira}, T.~M.~D., {De Pontieu}, B., \& {Carlsson}, M. 2012, \apj, 759, 18,
  \dodoi{10.1088/0004-637X/759/1/18}

\bibitem[{{Pereira} {et~al.}(2014){Pereira}, {De Pontieu}, {Carlsson},
  {Hansteen}, {Tarbell}, {Lemen}, {Title}, {Boerner}, {Hurlburt}, {W{\"u}lser},
  {Mart{\'{\i}}nez-Sykora}, {Kleint}, {Golub}, {McKillop}, {Reeves}, {Saar},
  {Testa}, {Tian}, {Jaeggli}, \& {Kankelborg}}]{2014ApJ...792L..15P}
{Pereira}, T.~M.~D., {De Pontieu}, B., {Carlsson}, M., {et~al.} 2014, \apjl,
  792, L15, \dodoi{10.1088/2041-8205/792/1/L15}

\bibitem[{{Pesnell} {et~al.}(2012){Pesnell}, {Thompson}, \&
  {Chamberlin}}]{2012SoPh..275....3P}
{Pesnell}, W.~D., {Thompson}, B.~J., \& {Chamberlin}, P.~C. 2012, \solphys,
  275, 3, \dodoi{10.1007/s11207-011-9841-3}

\bibitem[{{Pike} \& {Harrison}(1997)}]{1997SoPh..175..457P}
{Pike}, C.~D., \& {Harrison}, R.~A. 1997, \solphys, 175, 457,
  \dodoi{10.1023/A:1004987505422}

\bibitem[{{Pike} \& {Mason}(1998)}]{1998SoPh..182..333P}
{Pike}, C.~D., \& {Mason}, H.~E. 1998, \solphys, 182, 333,
  \dodoi{10.1023/A:1005065704108}

\bibitem[{{Popescu} {et~al.}(2007){Popescu}, {Xia}, {Banerjee}, \&
  {Doyle}}]{2007AdSpR..40.1021P}
{Popescu}, M.~D., {Xia}, L.~D., {Banerjee}, D., \& {Doyle}, J.~G. 2007,
  Advances in Space Research, 40, 1021, \dodoi{10.1016/j.asr.2007.06.068}

\bibitem[{{Raouafi} {et~al.}(2016){Raouafi}, {Patsourakos}, {Pariat}, {Young},
  {Sterling}, {Savcheva}, {Shimojo}, {Moreno-Insertis}, {DeVore}, {Archontis},
  {T{\"o}r{\"o}k}, {Mason}, {Curdt}, {Meyer}, {Dalmasse}, \&
  {Matsui}}]{2016SSRv..201....1R}
{Raouafi}, N.~E., {Patsourakos}, S., {Pariat}, E., {et~al.} 2016, \ssr, 201, 1,
  \dodoi{10.1007/s11214-016-0260-5}

\bibitem[{{Roberts}(1945)}]{1945ApJ...101..136R}
{Roberts}, W.~O. 1945, \apj, 101, 136, \dodoi{10.1086/144699}

\bibitem[{{Rouppe van der Voort} {et~al.}(2007){Rouppe van der Voort}, {De
  Pontieu}, {Hansteen}, {Carlsson}, \& {van Noort}}]{2007ApJ...660L.169R}
{Rouppe van der Voort}, L.~H.~M., {De Pontieu}, B., {Hansteen}, V.~H.,
  {Carlsson}, M., \& {van Noort}, M. 2007, \apjl, 660, L169,
  \dodoi{10.1086/518246}

\bibitem[{{Scherrer} {et~al.}(2012){Scherrer}, {Schou}, {Bush}, {Kosovichev},
  {Bogart}, {Hoeksema}, {Liu}, {Duvall}, {Zhao}, {Title}, {Schrijver},
  {Tarbell}, \& {Tomczyk}}]{2012SoPh..275..207S}
{Scherrer}, P.~H., {Schou}, J., {Bush}, R.~I., {et~al.} 2012, \solphys, 275,
  207, \dodoi{10.1007/s11207-011-9834-2}

\bibitem[{{Scullion} {et~al.}(2011){Scullion}, {Erd{\'e}lyi}, {Fedun}, \&
  {Doyle}}]{2011ApJ...743...14S}
{Scullion}, E., {Erd{\'e}lyi}, R., {Fedun}, V., \& {Doyle}, J.~G. 2011, \apj,
  743, 14, \dodoi{10.1088/0004-637X/743/1/14}

\bibitem[{{Scullion} {et~al.}(2009){Scullion}, {Popescu}, {Banerjee}, {Doyle},
  \& {Erd{\'e}lyi}}]{2009ApJ...704.1385S}
{Scullion}, E., {Popescu}, M.~D., {Banerjee}, D., {Doyle}, J.~G., \&
  {Erd{\'e}lyi}, R. 2009, \apj, 704, 1385, \dodoi{10.1088/0004-637X/704/2/1385}

\bibitem[{{Shibata}(1982)}]{1982SoPh...81....9S}
{Shibata}, K. 1982, \solphys, 81, 9, \dodoi{10.1007/BF00151974}

\bibitem[{{Shibata} {et~al.}(1992{\natexlab{a}}){Shibata}, {Nozawa}, \&
  {Matsumoto}}]{1992PASJ...44..265S}
{Shibata}, K., {Nozawa}, S., \& {Matsumoto}, R. 1992{\natexlab{a}}, \pasj, 44,
  265

\bibitem[{{Shibata} {et~al.}(1992{\natexlab{b}}){Shibata}, {Ishido}, {Acton},
  {Strong}, {Hirayama}, {Uchida}, {McAllister}, {Matsumoto}, {Tsuneta},
  {Shimizu}, {Hara}, {Sakurai}, {Ichimoto}, {Nishino}, \&
  {Ogawara}}]{1992PASJ...44L.173S}
{Shibata}, K., {Ishido}, Y., {Acton}, L.~W., {et~al.} 1992{\natexlab{b}},
  \pasj, 44, L173

\bibitem[{{Skogsrud} {et~al.}(2015){Skogsrud}, {Rouppe van der Voort}, {De
  Pontieu}, \& {Pereira}}]{2015ApJ...806..170S}
{Skogsrud}, H., {Rouppe van der Voort}, L., {De Pontieu}, B., \& {Pereira},
  T.~M.~D. 2015, \apj, 806, 170, \dodoi{10.1088/0004-637X/806/2/170}

\bibitem[{{Srivastava} \& {Murawski}(2011)}]{2011A&A...534A..62S}
{Srivastava}, A.~K., \& {Murawski}, K. 2011, \aap, 534, A62,
  \dodoi{10.1051/0004-6361/201117359}

\bibitem[{{Sterling}(2000)}]{2000SoPh..196...79S}
{Sterling}, A.~C. 2000, \solphys, 196, 79, \dodoi{10.1023/A:1005213923962}

\bibitem[{{Sterling} {et~al.}(2010){Sterling}, {Moore}, \&
  {DeForest}}]{2010ApJ...714L...1S}
{Sterling}, A.~C., {Moore}, R.~L., \& {DeForest}, C.~E. 2010, \apjl, 714, L1,
  \dodoi{10.1088/2041-8205/714/1/L1}

\bibitem[{{Suematsu}(2016)}]{2016ASPC..504..299S}
{Suematsu}, Y. 2016, in Astronomical Society of the Pacific Conference Series,
  Vol. 504, Coimbra Solar Physics Meeting: Ground-based Solar Observations in
  the Space Instrumentation Era, ed. I.~{Dorotovic}, C.~E. {Fischer}, \&
  M.~{Temmer}, 299

\bibitem[{{Suematsu} {et~al.}(1982){Suematsu}, {Shibata}, {Neshikawa}, \&
  {Kitai}}]{1982SoPh...75...99S}
{Suematsu}, Y., {Shibata}, K., {Neshikawa}, T., \& {Kitai}, R. 1982, \solphys,
  75, 99, \dodoi{10.1007/BF00153464}

\bibitem[{{Thompson} \& {Brekke}(2000)}]{2000SoPh..195...45T}
{Thompson}, W.~T., \& {Brekke}, P. 2000, \solphys, 195, 45,
  \dodoi{10.1023/A:1005203001242}

\bibitem[{{Waldmeier}(1955)}]{1955epds.book.....W}
{Waldmeier}, M. 1955, {Ergebnisse und Probleme der Sonnenforschung} (Leipzig:
  {Geest \& Portig})

\bibitem[{{Wang}(1998)}]{1998ApJ...509..461W}
{Wang}, H. 1998, \apj, 509, 461, \dodoi{10.1086/306497}

\bibitem[{{Wang} {et~al.}(2000){Wang}, {Li}, {Denker}, {Lee}, {Wang}, {Goode},
  {McAllister}, \& {Martin}}]{2000ApJ...530.1071W}
{Wang}, J., {Li}, W., {Denker}, C., {et~al.} 2000, \apj, 530, 1071,
  \dodoi{10.1086/308377}

\bibitem[{{Watanabe}(2014)}]{2014SPIE.9143E..1OW}
{Watanabe}, T. 2014, in \procspie, Vol. 9143, Space Telescopes and
  Instrumentation 2014: Optical, Infrared, and Millimeter Wave, 91431O

\bibitem[{{Wilhelm}(2000)}]{2000A&A...360..351W}
{Wilhelm}, K. 2000, \aap, 360, 351

\bibitem[{{Wilhelm} {et~al.}(1995){Wilhelm}, {Curdt}, {Marsch}, {Sch{\"u}hle},
  {Lemaire}, {Gabriel}, {Vial}, {Grewing}, {Huber}, {Jordan}, {Poland},
  {Thomas}, {K{\"u}hne}, {Timothy}, {Hassler}, \&
  {Siegmund}}]{1995SoPh..162..189W}
{Wilhelm}, K., {Curdt}, W., {Marsch}, E., {et~al.} 1995, \solphys, 162, 189,
  \dodoi{10.1007/BF00733430}

\bibitem[{{Withbroe}(1983)}]{1983ApJ...267..825W}
{Withbroe}, G.~L. 1983, \apj, 267, 825, \dodoi{10.1086/160917}

\bibitem[{{Withbroe} {et~al.}(1976){Withbroe}, {Jaffe}, {Foukal}, {Huber},
  {Noyes}, {Reeves}, {Schmahl}, {Timothy}, \& {Vernazza}}]{1976ApJ...203..528W}
{Withbroe}, G.~L., {Jaffe}, D.~T., {Foukal}, P.~V., {et~al.} 1976, \apj, 203,
  528, \dodoi{10.1086/154108}

\bibitem[{{Wuelser} {et~al.}(2004){Wuelser}, {Lemen}, {Tarbell}, {Wolfson},
  {Cannon}, {Carpenter}, {Duncan}, {Gradwohl}, {Meyer}, {Moore}, {Navarro},
  {Pearson}, {Rossi}, {Springer}, {Howard}, {Moses}, {Newmark},
  {Delaboudiniere}, {Artzner}, {Auchere}, {Bougnet}, {Bouyries}, {Bridou},
  {Clotaire}, {Colas}, {Delmotte}, {Jerome}, {Lamare}, {Mercier}, {Mullot},
  {Ravet}, {Song}, {Bothmer}, \& {Deutsch}}]{2004SPIE.5171..111W}
{Wuelser}, J.-P., {Lemen}, J.~R., {Tarbell}, T.~D., {et~al.} 2004, in
  \procspie, Vol. 5171, Telescopes and Instrumentation for Solar Astrophysics,
  ed. S.~{Fineschi} \& M.~A. {Gummin}, 111--122

\bibitem[{{Xia} {et~al.}(2005){Xia}, {Popescu}, {Doyle}, \&
  {Giannikakis}}]{2005A&A...438.1115X}
{Xia}, L.~D., {Popescu}, M.~D., {Doyle}, J.~G., \& {Giannikakis}, J. 2005,
  \aap, 438, 1115, \dodoi{10.1051/0004-6361:20042579}

\bibitem[{{Yamauchi} {et~al.}(2004){Yamauchi}, {Moore}, {Suess}, {Wang}, \&
  {Sakurai}}]{2004ApJ...605..511Y}
{Yamauchi}, Y., {Moore}, R.~L., {Suess}, S.~T., {Wang}, H., \& {Sakurai}, T.
  2004, \apj, 605, 511, \dodoi{10.1086/381240}

\bibitem[{{Yamauchi} {et~al.}(2005){Yamauchi}, {Wang}, {Jiang}, {Schwadron}, \&
  {Moore}}]{2005ApJ...629..572Y}
{Yamauchi}, Y., {Wang}, H., {Jiang}, Y., {Schwadron}, N., \& {Moore}, R.~L.
  2005, \apj, 629, 572, \dodoi{10.1086/431664}

\bibitem[{{Zhang} {et~al.}(2000){Zhang}, {Wang}, {Lee}, \&
  {Wang}}]{2000SoPh..194...59Z}
{Zhang}, J., {Wang}, J., {Lee}, C.-Y., \& {Wang}, H. 2000, \solphys, 194, 59,
  \dodoi{10.1023/A:1005287327897}

\bibitem[{{Zhang} \& {Ji}(2014)}]{2014A&A...561A.134Z}
{Zhang}, Q.~M., \& {Ji}, H.~S. 2014, \aap, 561, A134,
  \dodoi{10.1051/0004-6361/201322616}

\bibitem[{{Zhang} {et~al.}(2012){Zhang}, {Shibata}, {Wang}, {Mao}, {Matsumoto},
  {Liu}, \& {Su}}]{2012ApJ...750...16Z}
{Zhang}, Y.~Z., {Shibata}, K., {Wang}, J.~X., {et~al.} 2012, \apj, 750, 16,
  \dodoi{10.1088/0004-637X/750/1/16}

\end{thebibliography}

\appendix
\numberwithin{table}{section}
\renewcommand{\thetable}{\Alph{section}\arabic{table}}

\section{Supplementary data}

\let\storearraystretch\arraystretch
\def\arraystretch{1.0}

\begin{table*}[h!]
\centering
\caption{Distribution characteristics for the three groups of jets.}
\label{tbl:destr_par}
\begin{tabular}{lM{1.5cm}M{1.5cm}M{1.5cm}M{1.5cm}M{1.5cm}M{1.5cm}M{1.5cm}}
\hline\hline
& \multicolumn{7}{c}{All jets} \\
\cline{2-8}
& Minimum & Maximum & Mean & Median & Standard deviation & Gaussian fit mean & Gaussian fit sigma \\
\hline
Polar angle\tablenotemark{a} (deg) & -86.7 & 89.7 & 7.5 & 4.9 & 32.3 & 2.0 & 17.5 \\
Axis inclination (deg) & -40.4 & 41.3 & -2.1 & -1.5 & 16.2 & -0.9 & 15.4 \\
Maximum length (Mm) & 11.5 & 44.6 & 25.6 & 25.5 & 5.9 & 25.1 & 5.3 \\
Maximum width (Mm) & 2.3 & 12.0 & 4.6 & 4.4 & 1.3 & 4.4 & 1.3 \\
Length/width ratio & 2.4 & 13.6 & 6.0 & 5.7 & 1.9 & 5.5 & 1.6 \\
Lifetime (min) & 7.9 & 29.8 & 15.9 & 15.7 & 2.9 & 15.7 & 2.3 \\
Initial velocity (km\ps{}) & 50.2 & 199.3 & 109.0 & 109.8 & 23.2 & 109.1 & 24.4 \\
Deceleration (m\pss{}) & 56.1 & 563.5 & 240.1 & 236.6 & 80.3 & 227.8 & 72.6 \\
$E_\mathrm{kin}/E_\mathrm{pot}$ energy ratio & 0.21 & 2.39 & 0.91 & 0.89 & 0.31 & 0.87 & 0.28 \\
\hline
\end{tabular}
\vspace{8pt}\\
\hrulefill
\vspace{8pt}\\
\begin{tabular}{lM{1.5cm}M{1.5cm}M{1.5cm}M{1.5cm}M{1.5cm}M{1.5cm}M{1.5cm}}
\hline\hline
& \multicolumn{7}{c}{Coronal-hole jets} \\
\cline{2-8}
& Minimum & Maximum & Mean & Median & Standard deviation & Gaussian fit mean & Gaussian fit sigma \\
\hline
Polar angle\tablenotemark{a} (deg) & -40.6 & 37.2 & 1.9 & 2.3 & 13.4 & 2.1 & 15.0 \\
Axis inclination (deg) & -40.0 & 30.5 & -0.7 & -0.9 & 13.9 & -0.2 & 12.8 \\
Maximum length (Mm) & 16.1 & 44.6 & 27.5 & 26.7 & 5.1 & 26.5 & 3.6 \\
Maximum width (Mm) & 2.3 & 11.1 & 4.5 & 4.4 & 1.3 & 4.4 & 1.3 \\
Length / width ratio & 2.4 & 13.6 & 6.5 & 6.1 & 1.8 & 6.0 & 1.5 \\
Lifetime (min) & 10.4 & 29.8 & 16.5 & 16.2 & 2.6 & 16.1 & 2.0 \\
Initial velocity (km\ps{}) & 50.2 & 199.3 & 113.0 & 113.6 & 22.6 & 112.9 & 22.1 \\
Deceleration (m\pss{}) & 56.1 & 468.2 & 237.3 & 239.8 & 70.6 & 233.3 & 66.3 \\
$E_\mathrm{kin}/E_\mathrm{pot}$ energy ratio & 0.21 & 1.74 & 0.89 & 0.89 & 0.26 & 0.88 & 0.26 \\
\hline
\end{tabular}
\vspace{8pt}\\
\hrulefill
\vspace{8pt}\\
\begin{tabular}{lM{1.5cm}M{1.5cm}M{1.5cm}M{1.5cm}M{1.5cm}M{1.5cm}M{1.5cm}}
\hline\hline
& \multicolumn{7}{c}{Quiet-Sun jets} \\
\cline{2-8}
& Minimum & Maximum & Mean & Median & Standard deviation & Gaussian fit mean & Gaussian fit sigma \\
\hline
Polar angle\tablenotemark{a} (deg) & -86.7 & 89.7 & 17.2 & 26.2 & 49.0 & 58.1 & 8.2 \\
Axis inclination (deg) & -40.4 & 41.3 & -4.5 & -2.2 & 19.3 & -4.5 & 21.8 \\
Maximum length (Mm) & 11.5 & 42.6 & 22.4 & 20.7 & 5.9 & 20.5 & 4.5 \\
Maximum width (Mm) & 2.3 & 12.0 & 4.7 & 4.4 & 1.4 & 4.4 & 1.3 \\
Length / width ratio & 2.6 & 10.0 & 5.1 & 5.0 & 1.6 & 4.7 & 1.5 \\
Lifetime (min) & 7.9 & 25.2 & 14.9 & 14.9 & 3.2 & 14.5 & 2.8 \\
Initial velocity (km\ps{}) & 60.9 & 159.1 & 102.1 & 103.5 & 22.7 & 100.6 & 26.3 \\
Deceleration (m\pss{}) & 97.7 & 563.5 & 244.9 & 222.0 & 95.0 & 217.7 & 79.3 \\
$E_\mathrm{kin}/E_\mathrm{pot}$ energy ratio & 0.41 & 2.39 & 0.95 & 0.89 & 0.37 & 0.85 & 0.31 \\
\hline
\end{tabular}
\begin{flushleft}
\tablenotetext{a}{Polar angle is redefined here as relative to the nearest pole.}
\end{flushleft} 
\end{table*}

\begin{table*}[h!]
\centering
\caption{Pairwise Pearson's correlation coefficients of the jets' main characteristics.} \label{tbl:corr_par}
\begin{tabular}{lcccccc}
\hline\hline
& \multicolumn{6}{c}{All jets} \\
\cline{2-7}
& Polar angle\tablenotemark{a,b} & Axis inclination\tablenotemark{b} & Maximum length & Maximum width & Lifetime & Initial velocity \\
\hline
Deceleration & 0.13 & -0.02 & 0.13 & -0.10 & -0.77 & 0.81 \\
Initial velocity & -0.13 & 0.00 & 0.67 & 0.09 & -0.36 &  \\
Lifetime & -0.29 & 0.07 & 0.43 & 0.25 &  &  \\
Maximum width & 0.01 & 0.11 & 0.31 &  &  &  \\
Maximum length & -0.35 & 0.05 &  &  &  &  \\
Axis inclination\tablenotemark{b} & 0.09 &  &  &  &  &  \\
\hline
\end{tabular}
\vspace{8pt}\\
\hrulefill
\vspace{8pt}\\
\begin{tabular}{lcccccc}
\hline\hline
& \multicolumn{6}{c}{Coronal-hole jets} \\
\cline{2-7}
& Polar angle\tablenotemark{a,b} & Axis inclination\tablenotemark{b} & Maximum length & Maximum width & Lifetime & Initial velocity \\
\hline
Deceleration & 0.01 & -0.03 & 0.33 & -0.02 & -0.80 & 0.89 \\
Initial velocity & 0.09 & 0.05 & 0.71 & 0.16 & -0.53 &  \\
Lifetime & 0.06 & 0.09 & 0.18 & 0.21 &  &  \\
Maximum width & 0.23 & 0.13 & 0.38 &  &  &  \\
Maximum length & 0.16 & 0.14 &  &  &  &  \\
Axis inclination\tablenotemark{b} & 0.52 &  &  &  &  &  \\
\hline
\end{tabular}
\vspace{8pt}\\
\hrulefill
\vspace{8pt}\\
\begin{tabular}{lcccccc}
\hline\hline
& \multicolumn{6}{c}{Quiet-Sun jets} \\
\cline{2-7}
& Polar angle\tablenotemark{a,b} & Axis inclination\tablenotemark{b} & Maximum length & Maximum width & Lifetime & Initial velocity \\
\hline
Deceleration & 0.20 & -0.03 & -0.02 & -0.20 & -0.77 & 0.81 \\
Initial velocity & 0.04 & 0.09 & 0.55 & 0.03 & -0.35 &  \\
Lifetime & -0.25 & 0.20 & 0.57 & 0.37 &  &  \\
Maximum width & -0.22 & 0.05 & 0.37 &  &  &  \\
Maximum length & -0.20 & 0.23 &  &  &  &  \\
Axis inclination\tablenotemark{b} & 0.00 &  &  &  &  &  \\
\hline
\end{tabular}
\begin{flushleft}                           
\tablenotetext{a}{Polar angle is redefined here as relative to the nearest pole.}
\tablenotetext{b}{Polar angle and axis inclination are represented here by the absolute values, except for their mutual correlation.}
\end{flushleft} 
\end{table*}

\let\arraystretch\storearraystretch

\end{document}